%% file: main.tex
\documentclass[twocolumn]{aastex62}

\newcommand{\todo}[1]{\textcolor{red}{\textbf{#1}}}

\newcommand{\arcs}{$^{\prime\prime}$} % Arcseconds
\newcommand{\beq}{\begin{equation}\begin{aligned}}
\newcommand{\eeq}{\end{aligned}\end{equation}}

\newcommand{\msun}{M$_\odot$}

\usepackage{scalerel,amssymb}
\usepackage{amsmath}
\usepackage[caption=false]{subfig}

\usepackage[]{url}
\graphicspath{{./}{figures/}}

\accepted{\today}
\submitjournal{ApJ}

\shorttitle{Dwarf satellite systems in the Local Volume.}
\shortauthors{Carlsten et al.}

\begin{document}

\title{Wide-Field Survey of Dwarf Satellite Systems Around 10 Hosts in the Local Volume}

\correspondingauthor{Scott G. Carlsten}
\email{scottgc@princeton.edu}

\author[0000-0002-5382-2898]{Scott G. Carlsten}
\affil{Department of Astrophysical Sciences, 4 Ivy Lane, Princeton University, Princeton, NJ 08544}

\author[0000-0003-4970-2874]{Johnny P. Greco}
\altaffiliation{NSF Astronomy \& Astrophysics Postdoctoral Fellow}
\affiliation{Center for Cosmology and AstroParticle Physics (CCAPP), The Ohio State University, Columbus, OH 43210, USA}

\author[0000-0002-1691-8217]{Rachael L. Beaton}
\altaffiliation{Hubble Fellow}
\affiliation{Department of Astrophysical Sciences, 4 Ivy Lane, Princeton University, Princeton, NJ 08544}
\affiliation{The Observatories of the Carnegie Institution for Science, 813 Santa Barbara St., Pasadena, CA~91101\\}

\author{Jenny E. Greene}
\affil{Department of Astrophysical Sciences, 4 Ivy Lane, Princeton University, Princeton, NJ 08544}

\begin{abstract}
We present the results of an extensive search for dwarf satellite galaxies around 10 primary host galaxies in the Local Volume (D$<$12 Mpc) using archival CFHT/MegaCam imaging data. The hosts span a wide range in properties, with stellar masses ranging from that of the LMC to ${\sim}3$ times that of the Milky Way (MW). The surveyed hosts are: NGC 1023, NGC 1156, NGC 2903, NGC 4258, NGC 4565, NGC 4631, NGC 5023, M51, M64, and M104. We detect satellite candidates using a consistent semi-automated detection algorithm that is optimized for the detection of low surface brightness objects. Depending on the host, our completeness limit is $M_g{\sim}-8$ to $-10$ (assuming the distance of the host). We detect objects with surface brightness down to $\mu_{0,g}{\sim}26$ mag arcsec$^{-2}$ at $\gtrsim90\%$ completeness. The survey areas of the six best-surveyed hosts cover most of the inner projected $R<150$ kpc area, which roughly doubles the number of MW-mass hosts surveyed at this level of area and luminosity completeness. The number of detected candidates range from 1 around M64 to 33 around NGC 4258. In total, 153 candidates are found, of which 93 are new. While we defer an analysis of the satellite luminosity functions of the hosts until distance information is available for the candidates, we do show that the candidates are primarily red, spheroid systems with properties roughly consistent with known satellites in the Local Group.
\end{abstract}

\section{Introduction}
Discovering nearby galaxies of ever lower mass is a continual goal of extragalactic astronomy. A full census of galactic systems is required to robustly test structure formation theories on the smallest of scales. In recent years, these searches are often motivated by the well-known `small-scale challenges' to $\Lambda$CDM \citep[e.g.][]{bullock2017}, including the Missing Satellite Problem \citep[e.g.][]{moore1999, klypin1999}, the Too Big to Fail problem \citep[e.g.][]{bk2011,bk2012}, and the Planes of Satellites Problem \citep[e.g.][]{pawlowski2012,ibataGPOA,muller_plane}. 

The search for low mass galaxies that are satellites of higher mass hosts has been facilitated by modern, wide-field survey imaging. Surveys like SDSS, DES, Pan-STARRS, and HSC-SSP have revealed ${\sim}$50 satellite companions around the Milky Way (MW) \citep[e.g.][]{belokurov2008,belokurov2010,laevens2015,drlica2015,bechtol2015,homma2016,homma2018,homma2019,koposov2018,kim2015}. In these cases, dwarfs are found as overdensities in star counts. The distance to the dwarfs can then be found directly from the resolved color-magnitude diagram. A similar approach can be taken with other hosts that are near enough (D$\lesssim$3-4 Mpc) for RGB stars to be resolved and detected from the ground. Much of M31's halo has been surveyed by the PANDAS project \citep{mcconnachie2009,martin2016, mcconnachie2018} using CFHT/MegaCam, revealing $>$20 dwarf satellite companions. M81 has been similarly surveyed with CFHT/MegaCam \citep{chiboucas2009,chiboucas2013}. Centaurus A has been surveyed by multiple groups \citep{taylor2016,mullerCenA,muller2019,crnojevic2019}. Finally, both M94 and NGC 2403 have been surveyed with Subaru/HSC \citep{smercina2018,madcash,carlin2019}. Surveying nearby systems like these has the significant drawback that the virial volume of these hosts subtends a large area on the sky and requires large investment of telescope time to completely cover. 

Studying hosts in the D${\sim}$5-10 Mpc range would help significantly to reduce the amount of sky coverage needed. The dwarfs at these distances are detected from integrated light as resolved stars are not detected from the ground. Because of this, distances to the dwarfs become the significant roadblock to these searches. From the survey images alone, it is often not clear whether the dwarfs are genuine companions or background (or even foreground) contaminants. Contamination fractions can be quite high \citep[$>$80\%,][]{sbf_m101,cohen2018,bennet2019}. Therefore, often spectroscopic or \emph{HST} follow-up is required, increasing the telescope investment. \emph{HST} follow-up must be undertaken in a source-by-source fashion due to the small FOV of \emph{HST}.

With deep enough photometric data, the distances to low surface brightness (LSB) dwarfs can be measured directly from the images in the D${\sim}$5-10 Mpc range using surface brightness fluctuation (SBF) measurements \citep{tonry1988, tonry2001, jerjen_field, jerjen_fornax}. \citet{sbf_calib} explored this technique and showed that distances accurate to ${\sim}$15\% are possible, which will often be enough to confirm a satellite's association with a host and provided a calibration based completely on TRGB distances. \citet{sbf_m101} applied this to the catalog of candidate satellites of M101 of \citet{bennet2017} and was able to show that the majority were background contaminants while confirming two as actual companions. The results of \citet{sbf_m101} have since been confirmed by \citet{bennet2019} using considerable \emph{HST} follow-up. 

Including M101, there are now 6 (MW, M31, M81, M94, Cen A, and M101) nearby MW-sized hosts that have been surveyed out to a large fraction of the host's virial radius\footnote{Note that the MW's satellite census is complicated due to the limited sky coverage of existing surveys and the zone of the avoidance of the disk \citep{newton2018}.} with a high level of completeness. These surveys are sensitive to dwarfs with luminosities $M_V \lesssim -10$. This sample is sufficiently large that the scatter in satellite systems among hosts is starting to be explored \citep[e.g.][]{smercina2018}, in addition to how the properties of satellites correlates with those of the host \citep[e.g.][]{bennet2019}. However, the sample is still too small for rigorous statistical tests. Other systems have been surveyed \citep[e.g.][]{trentham2009, kim2011, park2017, cohen2018, mullerLeo, tanaka2018} but either the lack of distance information and/or the lack of completeness estimates make interpretation difficult. 

A complementary approach to the searches described above is being taken by the SAGA Survey \citep{geha2017}. The SAGA Survey uses extensive spectroscopic follow-up of almost all SDSS detected galaxies around MW-sized hosts within $20<D<40$ Mpc to confirm satellite membership via line-of-sight velocity. A major draw-back is that using SDSS detected galaxies for targeting means that only satellites with $M_r<-12.3$ (at $D=20$ Mpc) will be included. Thus far, 8 MW analogs have been surveyed out to their virial radii. The number of satellites in this luminosity range per host ranges from 1 to 9, indicating large host-to-host scatter. A striking first result is that 26 out of the 27 detected satellites are actively forming stars (based on H$\alpha$ emission in the spectra), and the majority have irregular morphologies. This is in stark contrast to the MW satellite system, which is dominated by quenched spheroidal systems. Only 2 out of 5 MW satellites in this luminosity range are actively forming stars. 
The SAGA Survey will provide excellent statistics on the occurrence and properties of bright satellites, but detecting very faint satellites is only possible in the LV which is crucial to learn about galaxy formation in the smallest dark matter halos and the impact that reionization has on these galaxies \citep[e.g.,][]{jethwa2018, bose2018, kim2018, nadler2019}. With that said, the very faintest satellites will likely only ever be discoverable and characterized around the MW, but we will only ever have one data point. Studies of satellites systems in the LV represent a sort of sweet spot where many faint satellites can be discovered around a single host, and multiple hosts can still be surveyed.

\begin{deluxetable*}{cccccccccc}
\tablecaption{Host Properties\label{tab:sample}}

\tablehead{\colhead{Name} & \colhead{$\alpha$} & \colhead{$\delta$} & \colhead{Dist} & \colhead{$V_{\rm circ}$} & \colhead{M$_V$} & \colhead{M$_{K_s}$} & \colhead{$cz$} & \colhead{Median Depth} & \colhead{Filters}  \\ 
\colhead{} & \colhead{} & \colhead{} & \colhead{(Mpc)} & \colhead{(km/s)} & \colhead{(mag)} & \colhead{(mag)} & \colhead{(km/s)} & \colhead{(s)} & \colhead{} } 
\startdata
 NGC 1023 & 02:40:24 & +39:03:48 &   10.4$^{i}$ &  250$^{p}$   & -20.7$^{a}$ & -23.8$^{r}$ & 638$^{s}$ & 1640/11800 & g/i   \\
 NGC 1156 & 02:59:43 & +25:14:28  &   7.6$^{j}$ &   55$^{b}$  & -17.7$^{a}$ & -19.9$^{r}$ & 379$^{s}$ & 7700/3400 & g/r  \\
 NGC 2903 & 09:32:10 & +21:30:03 &   8.0$^{i}$ &   189$^{b}$   & -20.4$^{c}$ & -23.5$^{r}$ & 556$^{s}$ & 940/1400 & g/r     \\
 NGC 4258 & 12:18:58 & +47:18:13 &  7.2$^{k}$ &   208$^{d}$   & -20.9$^{a}$ & -23.8$^{r}$ &462$^{s}$ & 840/2580 & g/r  \\
 NGC 4565 & 12:36:21 & +25:59:15 &   11.9$^{l}$   & 244 &   -17.9$^{e}$ & -24.3$^{r}$ &1261$^{s}$ & 3360/4400 & g/r\\
 NGC 4631 & 12:42:08 & +32:32:29 &   7.4$^{l}$   & 127$^{b}$   & -20.2$^{a}$ & -22.9$^{r}$ & 606$^{s}$ & 1700/3660 & g/r   \\
 NGC 5023 & 13:12:12 & +44:02:17 &   6.5$^{l}$   & 78$^{b}$   & -14.9$^{q}$ & -19.3$^{r}$ & 404$^{s}$ &  1750/2250 & g/i  \\
 M51 &   13:29:53 & +47:11:43 &   8.6$^{m}$  & 220$^{f}$   & -21.3$^{a}$ & -24.2$^{r}$ & 465$^{s}$& 2450/2530 & g/r  \\
 M64 &   12:56:44 & +21:40:58 &   5.3$^{n}$   & 155$^{g}$   & -20.1$^{a}$ & -23.3$^{r}$ & 402$^{s}$& 1400/2370 & g/r   \\
 M104 &   12:39:59 & -11:37:23 &   9.55$^{o}$   & 380$^{h}$   & -21.9$^{a}$ & -24.9$^{r}$ & 1092$^{s}$& 3000/3200 & g/i    \\
\enddata
\tablecomments{Properties of the hosts surveyed in this work. Median depth throughout the surveyed area is given for both filters used: $g$ and either $r$ or $i$.}
\tablerefs{$^{a}$ - \citet{galex2007}
$^{b}$ - \citet{karachentsev}
$^{c}$ - \citet{lvl2014}
$^{d}$ - \citet{m106_vcirc}
$^{e}$ - \citet{quasar_cat}
$^{f}$ - \citet{m51_vcirc1}
$^{g}$ - \citet{rubin1994}
$^{h}$ - \citet{m104_vcirc}
$^{i}$ - NED Median
$^{j}$ - \citet{kim2012}
$^{k}$ - \citet{m106_maser}
$^{l}$ - \citet{ngc4565_dist}
$^{m}$ - \citet{m51_dist}
$^{n}$ - \citet{m64_dist}
$^{o}$ - \citet{m104_dist}
$^{p}$ - \citet{ngc1023_vcirc}
$^{q}$ - \citet{sdss7} 
$^{r}$ - \citet{2mass}
$^{s}$ - SIMBAD
}

\end{deluxetable*}

Another motivation for studying the satellite systems of nearby galaxies is the discovery of two systems with apparently anomalously very low dark matter content, NGC 1052-DF2 and DF4 \citep{pvd2018,danieli_df2,pvd_df4}. Both galaxies were discovered as LSB satellites of NGC 1052 \citep[see, however,][for a different interpretation]{trujillo2019, df4_distance}. Finding other examples of these galaxies would confirm their existence and give insights to their formation \citep[e.g.][]{ogiya2018, silk2019}. Looking for large, LSB dwarf satellites that show a significant collection of associated point sources seems an excellent place to start \citep[e.g.][]{forbes2019}.

To increase the sample of well-characterized satellite systems, we have carried out a search using archival CFHT/MegaCam imaging data for satellites around a wide variety of hosts in the LV (D$\lesssim$12 Mpc). Since its first-light in 2003, MegaCam \citep{megacam} has accumulated an extensive archive\footnote{\url{http://www.cadc-ccda.hia-iha.nrc-cnrc.gc.ca/en/cfht/}} of imaging data, including many nearby, massive primary galaxies. In this paper, we search through the imaging data of 10 such hosts for satellite companions in a homogeneous, semi-automated way, carefully quantifying our completeness. We plan to confirm distances with SBF measurements, where possible, in a future paper.

Our sample of primaries includes hosts that are both more and less massive than the MW. Host galaxies with different masses will interact with their satellites in different ways. For instance, low mass hosts are not expected to have hot gas halos \citep{birnboim2003} and the baryonic disks of these galaxies are expected to less severely tidally disrupt subhalos due to a higher mass-to-light ratio compared to MW-size hosts \citep{chan2019}. Characterizing the satellites of these low mass hosts will help unravel what are the important physical processes that determine the properties of the satellites \citep{carlin2019}. Additionally, two galaxies in our sample are field Large Magellanic Cloud (LMC) analogs, and their satellite systems will help put the inferred satellite system of the LMC \citep[e.g.][]{sales2017, Kallivayalil2018, pardy2019} in context. Of the confirmed satellites of the LMC, there is a notable dearth of large $M_\star>10^4$ \msun~satellites compared to theoretical predictions \citep{dooley2017}\footnote{Confirming Carina and Fornax as satellites \citep{pardy2019} will alleviate this tension.}. \citet{dooley2017} predict 1-6 dwarfs with $M_\star>10^5$ \msun~in the virial volume of field LMC analogs. While  $M_\star{\sim}10^5$ \msun~dwarfs will generally be below our completeness limit,  $M_\star{\sim}10^6$ \msun~dwarfs will be easily detectable. Determining the average number of systems of this mass around LMC analogs can help put constraints on the stellar-to-halo mass relation \citep{dooley2017}. Studying the satellites of more massive hosts will show how satellite abundance and properties scale with host mass, like has been done with SDSS \citep{wenting2012, sales2013} but at far lower satellite luminosities and higher completeness.

In Section \ref{sec:data} we describe the sample selection, properties of the hosts, and the data reduction. In Section \ref{sec:detection}, we describe the semi-automated dwarf detection algorithm used. In Section \ref{sec:completeness}, we describe the completeness tests. In Section \ref{sec:catalogs}, we present our catalogs of discovered dwarfs. Finally, in Section \ref{sec:discussion}, we discuss the resulting catalogs of satellites and conclude in Section \ref{sec:conclusion}.

\section{Data \& Sample}
\label{sec:data}
\subsection{Sample Selection}
We start the selection of hosts to survey with the Revised Shapley-Ames Catalog of Bright Galaxies \citep{shapley-ames}. We first select galaxies with redshifts $cz < 1500$ km/s and then search for these in the CFHT data archive. We require the galaxy to have imaging in either $g$ and $r$ or $g$ and $i$. This leaves ${\sim}70$ hosts. We select the 10 galaxies that have the longest exposure times and closest distances as the highest priority targets to search. Each host has at least ${\sim}1$ sq. degree of surveyed area as that is the FOV of MegaCam, but several of the hosts have multiple pointings. The sample is described in Table \ref{tab:sample}. These galaxies are all within $D<$12 Mpc but otherwise span a range in properties.

\begin{figure*}
\includegraphics[width=\textwidth]{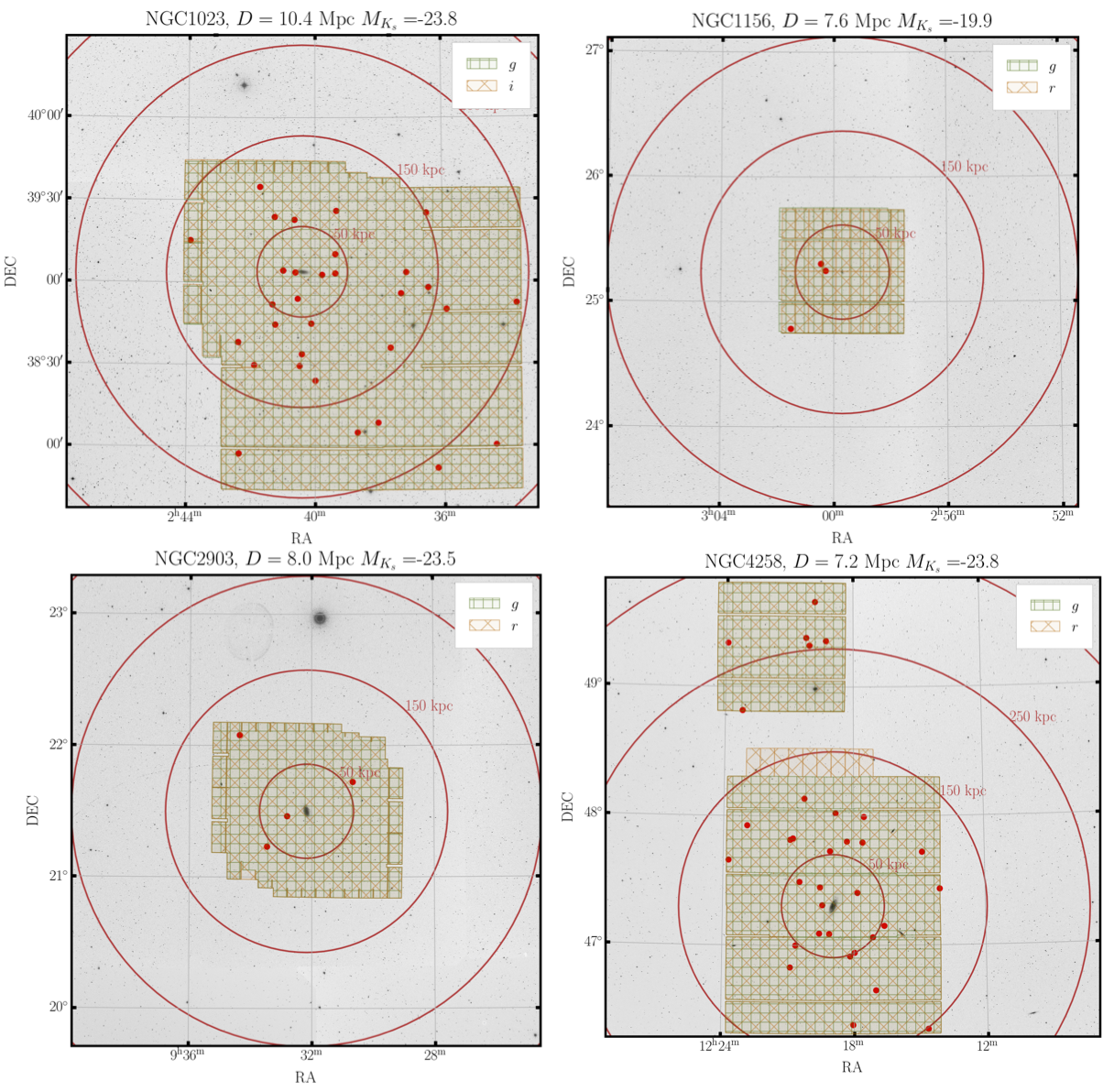}
\caption{The surveyed area of NGC 1023, NGC 1156, NGC 2903, and NGC 4258. The footprints in each of the two bands used for the hosts are shown in the hatched area. The background image is from DSS. Radii of different physical size at the distance of the host are shown. The red points show the locations of the candidate satellites detected in this work.}
\label{fig:area1}
\end{figure*}

\begin{figure*}
\ContinuedFloat
\includegraphics[width=\textwidth]{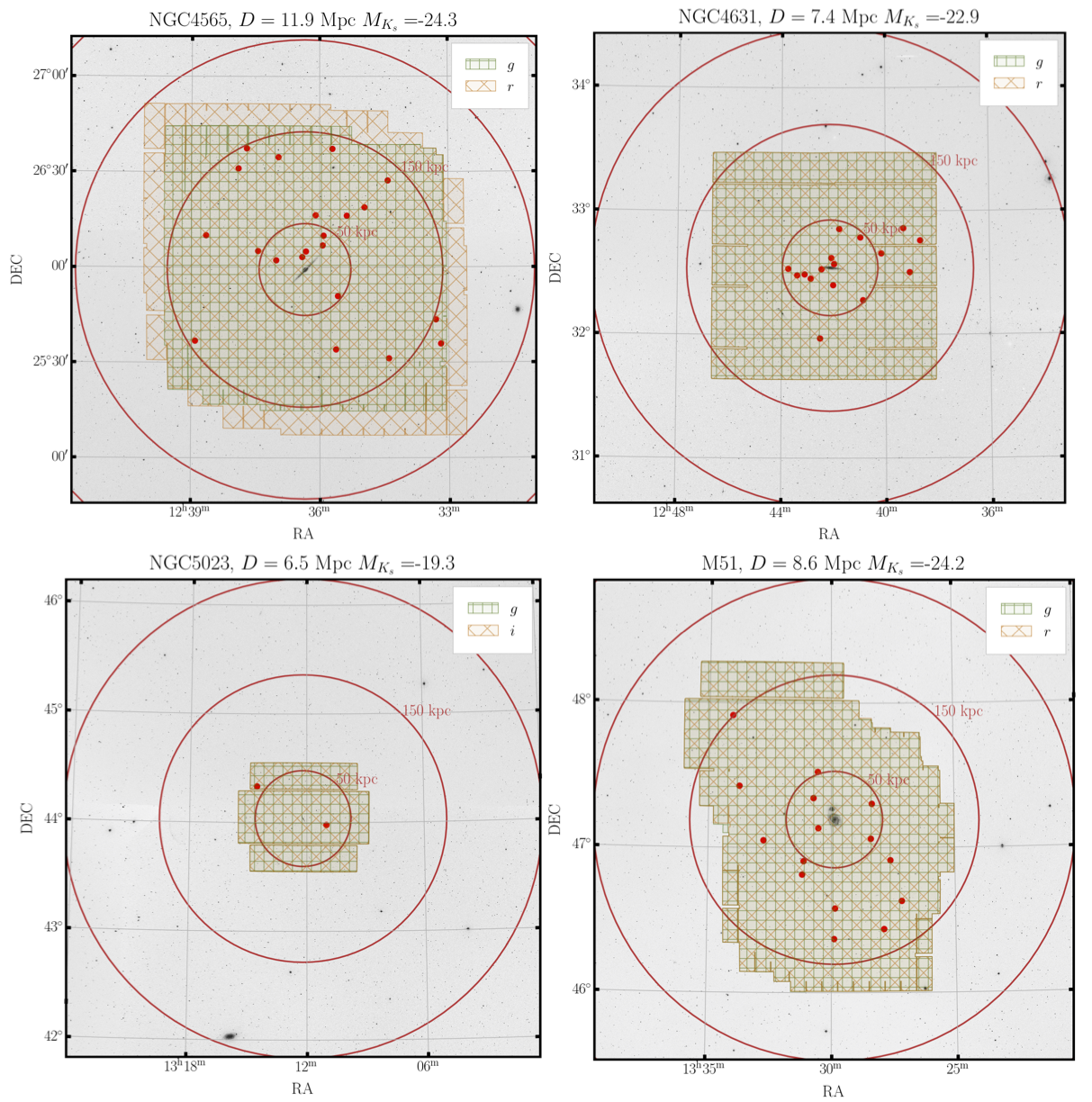}
\caption{The surveyed area of NGC 4565, NGC 4631, NGC 5023, and M51.}
\label{fig:area2}
\end{figure*}

\begin{figure*}
\ContinuedFloat
\includegraphics[width=\textwidth]{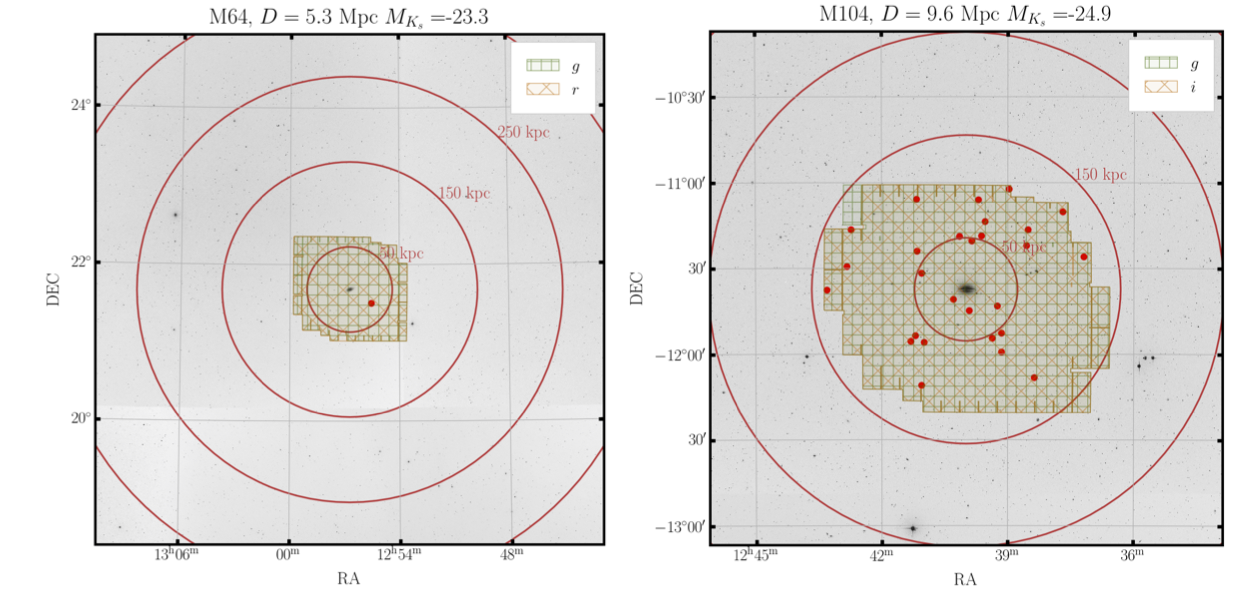}
\caption{The surveyed area of M64 and M104.}
\label{fig:area3}
\end{figure*}

Figure \ref{fig:area1} shows the survey footprints for each of the ten host galaxies. As expected, the coverage is heterogeneous, but for six of the hosts (M104, NGC 4258, NGC 1023, NGC 4631, M51, NGC 4565) a large fraction ($\gtrsim75$\%) of the inner projected 150 kpc area is covered\footnote{An even larger fraction of the inner 150 kpc \textit{volume} will be covered.}. For the two lowest mass hosts (NGC 1156 and NGC 5023), the coverage is limited to the inner ${\sim}$50 kpc but the virial radii of these systems are ${\sim}$100kpc \citep{madcash}, so much of the virial volume is still surveyed. The coverage for M64 and NGC 2903 is less but we include them in our analysis nonetheless.

\subsection{Overview of Host Galaxies}
In this section, we provide an overview of the host sample. Detailed information for each host is presented in \S\ref{sec:catalogs} when discussing the catalogs of candidate satellites, but we provide a brief overview of the range in host properties in this section. Most significantly, our hosts range in mass from two LMC-mass analogs (NGC 1156 and NGC 5023) to M104 which is several times the mass of the MW (in stellar and total). Additionally, the sample spans a range in environment from the group environment of NGC 1023 to the isolated M64 and NGC 1156.  Several of our hosts have been explicitly searched for satellites before. In particular, NGC 1023, NGC 4258, and NGC 4631 have been surveyed with deep and wide-field imaging. We revisit these as well, to have a uniformly selected and completeness-tested sample on which to compare the satellite systems. Also it is useful to compare our lists with prior work. Some previous searches were done with small telescopes, but some were done with the same MegaCam data that we use in the current survey.

\subsection{Data Reduction}
For each of the ten hosts detailed above, we download all available imaging data in the two bands used from the CADC archive\footnote{\url{http://www.cadc-ccda.hia-iha.nrc-cnrc.gc.ca/en/cfht/}}. We start with the \texttt{Elixir} \citep{elixir} pre-processed CCD frames and perform the calibration, sky subtraction, and stacking ourselves as outlined in \citet{sbf_calib}. The \texttt{Elixir} pre-processed images have had the instrumental signatures removed and have been flat-fielded. The images are also given a rough, starting astrometric and photometric calibration. We improve the astrometric calibration by cross-referencing each chip, using the \texttt{Scamp} \citep{scamp} software, with SDSS-DR9 \citep{sdss_dr9} sources or USNO-B1 \citep{usnob} for galaxies outside of the SDSS footprint. The photometric calibration is done by cross-referencing stars on the chips with sources from SDSS-DR14 \citep{sdss_df14} or Pan-STARRS1 \citep{panstarrs, panstarrs2}. SDSS and Pan-STARRS1 magnitudes are converted to the MegaCam photometric system using transformation equations\footnote{Available online \url{http://www.cadc-ccda.hia-iha.nrc-cnrc.gc.ca/en/megapipe/docs/filt.html}}. NGC 1156 is outside of the SDSS footprint and, unfortunately, falls in one of the holes in the Pan-STARRS-DR2 footprint\footnote{\url{https://outerspace.stsci.edu/display/PANSTARRS/PS1+DR2+caveats}}. For this galaxy, we do the photometric calibration via a cross-match with Gaia-DR2 \citep{gaia_dr2} stars. We use the photometry in the NGC 4258 field (which is calibrated with an SDSS cross-match) to generate these empirical $G^{\rm Gaia}$ to $g^{\rm CFHT}$ and $r^{\rm CFHT}$ conversions as functions of the $BP-RP$ color:
\beq
G-r = -0.255x^2 + 0.482x -0.141\\
G-g = -0.028x^2 - 0.762x + 0.312
\eeq
where $x\equiv BP-RP$.
Since the \textit{Gaia} passbands are much wider than the MegaCam bands, the scatter in this conversion is larger than that of the SDSS or Pan-STARRS1 to MegaCam conversions at around \todo{~}0.2 mag. We include this in the uncertainty in the photometry reported for this host.

Once the chip images are calibrated, they are resampled onto a common pixel grid with \texttt{Swarp} \citep{swarp} and then median-coadded. We allow for \texttt{Swarp} to do a local background subtraction using a mesh size of 256$\times$256 pixels. In \citet{sbf_calib}, we found that such a local subtraction was inadequate for SBF measurements. However, with these stacks, we will only intend to detect the dwarfs. For photometry of the dwarfs (and later SBF measurements), we re-reduce the area around each dwarf as described below. The images are coadded into 8k$\times$8k pixel tracts for convenience in the dwarf detection step. The number of tracts range from 9, for galaxies with only one MegaCam pointing, to 34 for NGC 4258 with several pointings. 

All reported photometry are AB magnitudes in the CFHT/MegaCam photometric system, unless otherwise stated. Photometry is corrected for Galactic extinction using the $E(B-V)$ values from \citet{sfd} recalibrated by \citet{sfd2}.

\section{Dwarf Galaxy Detection}
\label{sec:detection}
While much previous work in this field has used visual searches \citep[e.g.][]{kim2011, park2017, smercina2018, crnojevic2019, mullerLeo, muller101, mullerCenA}, we opt to do a semi-automated search in the vein of \citet{bennet2017} and \citet{johnny2}, due to the large combined area in our present survey and the need to do extensive completeness checks. In short, an automated detection step selects candidate dwarf galaxies which are then visually inspected for confirmation. 

Our goal is not just to characterize the LSB satellites for nearby hosts but also any higher surface brightness dwarf satellites, if present. Therefore we do two detection steps: one optimized for larger higher surface brightness systems and one optimized for the LSB dwarfs. This ``hot and cold'' two-step detection process is often used when detection of both HSB and LSB systems is necessary \citep[e.g.][]{rix2004, leauthaud2007, prescott2012}.

For reference in the following discussion, the scaling relations of \citet{danieli_field} indicate that a $10^8$\msun~ dwarf will have an average SB within the effective radius of $\mu_g{\sim}22$ to $22.5$ mag arcsec$^{-2}$\footnote{Depending on the color (since we convert from $V$ to $g$).} and angular size of $\sim$20\arcs~to 40\arcs~ for the distances of our hosts. Similarly, a $10^5$\msun~ dwarf will have an average SB within the effective radius of $\mu_g{\sim}28$ mag arcsec$^{-2}$ and angular size of $\sim$3\arcs~to 6\arcs.

The major steps in the detection process are:
\begin{enumerate}
  \item Generate and apply star masks to the images using \textit{Gaia} star catalogs.
  \item Detect large, HSB ($\mu_0\lesssim24$ mag arcsec$^{-2}$) satellite candidates.
  \item Mask bright background sources and associated diffuse light. Sources $\gtrsim15\sigma$ above the background are masked and these masks are effectively grown to include associated diffuse light down to ${\sim}1\sigma$ above the background. Mask remaining sources down to ${\sim}4\sigma$ above the background.
  \item Filter the image with a Gaussian with FWHM ${\sim}1\times$ the FWHM of the PSF. Detect sources on the filtered, masked image.
  \item Visually inspect detections to remove false positives from detected background galaxies.
\end{enumerate}
In the following subsections, we go through each step in more detail. Each step is done independently in each of the two bands ($g/r$ or $g/i$) that we use for the regions. Detections are merged between the two filters and then detections are visually inspected for confirmation, as described below.  The main steps of the masking and LSB object detection  are shown in Figure \ref{fig:example_det}.

\subsection{Step 1: Star Masks}
The first step is to mask bright stars and the halos of scattered light surrounding them. This is especially crucial because MegaCam exhibits large (${\sim}8^{\prime}$ wide) and very prominent donut-shaped scattered light halos around very bright stars \citep[e.g.][]{atlas3d}. The size of mask that each star needs depends on its brightness. To make the star masks, we start with a catalog of stars in each region from \textit{Gaia} DR2 \citep{gaia_dr2}. We use \textit{Gaia} magnitudes since the stars will almost all be saturated in the MegaCam data. For each region, we detect stars in the MegaCam data with \texttt{SExtractor} \citep{SExtractor} and derive a relation between the stars' \textit{Gaia} magnitude and the \texttt{SExtractor} \texttt{ISOAREA\_IMAGE}, which represents the size that the mask needs to be for that star. This relation is scaled so that the masks are grown to include LSB scattered light that was below the \texttt{SExtractor} detection threshold. Then, the location and magnitude of all the stars in the \textit{Gaia} catalog are used to construct a mask for each region. 

For stars brighter than ${\sim}9^{th}$ magnitude (the exact limit depends on the exposure time), a large prominent halo becomes visible in the MegaCam imaging. The halos are more pronounced in the redder bands, although they are present in all bands. These halos would not be covered by the star mask described above and do not get bigger for brighter stars (they do get more pronounced, however). We therefore identify these stars from their \textit{Gaia} magnitude converted into Sloan-$i$, using a threshold of $i<9$, and mask a 2600 pixel (${\sim}8^{\prime}$) wide circle around each of them. 

For most of the galaxies, these star masks constitute a small ($\lesssim$5\%) fraction of the survey area. Due to its low galactic latitude, NGC 1023 is the most significantly affected by this with roughly 12\% of the area getting lost to these masks. While these large halos would be disastrous for any automated LSB galaxy detection, it is often still possible to see and identify by-eye dwarf galaxies that fall in them. Therefore, we reserve these cutouts for the visual inspection step described below. While this constitutes a fairly large fraction of the NGC 1023 surveyed area, it is much less of the area in the other fields. The final dwarf catalogs would not change significantly if these masked regions were completely thrown out instead. We keep track of which dwarfs were found by eye in these masked areas making it possible to check whether any science conclusions change if the dwarfs found by eye in the scattered light halos are omitted. We note that only the masked regions around the brightest stars ($\lesssim9^{th}$ mag) are inspected by eye; the star masks for the dimmer stars that do not exhibit the large halos of scattered light are not.

All pixels that are masked in this and further masking steps are replaced by sky noise. The background and rms used to estimate the sky level/noise are derived with \texttt{SExtractor} using a meshsize of 128$\times$128. The images have already been background subtracted at this stage, so the background levels found here are very close to zero.

\subsection{Step 2: HSB Object Detection}
With the stars and their associated scattered light masked, we first detect large, HSB candidate satellites. When detecting the HSB dwarf satellite companions, we primary distinguish between possible companions and bright background galaxies based on size. The scaling relations of LG dwarfs \citep[e.g.][]{danieli_field} indicate that HSB dwarfs will generally be quite large. Selecting for HSB candidates with this prior will mean we are insensitive to compact systems, like M32, but there is little evidence to suspect that systems as compact as M32 are common \citep[cf. Sec 5.3 of ][]{geha2017}. Comparable searches to ours are also insensitive to these types of systems \citep[e.g.][]{bennet2017, geha2017}\footnote{The SAGA Survey is biased against these compact systems because they select SDSS sources that are resolved for spectroscopic follow-up.}. We will also be insensitive to ultra compact dwarf (UCD) satellites. UCDs are most commonly found in higher mass cluster-like environments, which are quite different to the environments of our hosts. With that said, a UCD is known in the M104 system \citep{m104_ucd}, and a candidate is known in the NGC 1023 system \citep{ngc1023_ucd}. Even if UCDs were present in our target systems, it is unclear whether they should really be counted as satellites, as it depends on how they are formed \citep{mieske_ucd}. 

We detect HSB dwarfs with \texttt{SExtractor} using a threshold of 5$\sigma$ above the background and a minimum detected area of 1200 pixels (corresponding to a circle with radius ${\sim}4$\arcs~or ${\sim}150$pc at 8 Mpc) above the detection threshold. For the regions with the deepest data (NGC 1023, NGC 4258, and NGC 4565), we increased the minimum area to 1800 pixels to reduce contamination by HSB background galaxies. After detection, the sources are restricted to those with average surface brightness within the effective radius (approximated by \texttt{SExtractor}'s \texttt{FLUX\_RADIUS}) fainter than 22.5 mag arcsec$^{-2}$. This size and surface brightness cut for the HSB objects were chosen to minimize contaminants from background galaxies while still recovering the HSB companions that have been discovered in previous searches.

 \begin{figure*}
\includegraphics[width=\textwidth]{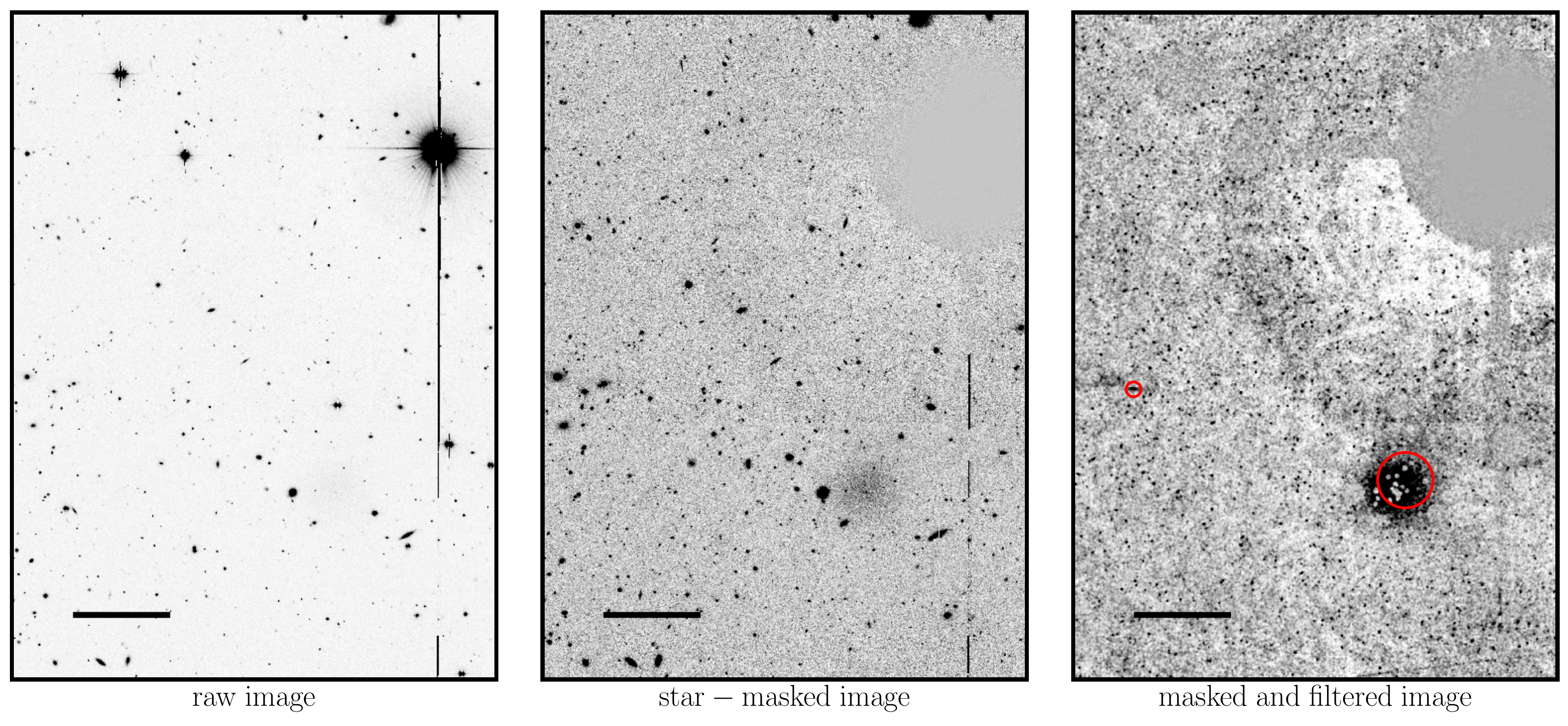}
\caption{A demonstration of the LSB galaxy detection algorithm. The leftmost panel shows a cutout of the coadded data around M104. The middle panel shows the cutout once the star mask has been applied. The rightmost panel shows the cutout once the low/high threshold masking and point source masking has been completed and the image has been filtered to bring out LSB diffuse light. The red circles show the detected objects. The larger detected object is a high priority dwarf candidate while the smaller is a blend of background galaxies that gets removed in the visual inspection step. The detection of the dwarf is off-center because some of the brightest central parts get masked. On the right, a faint halo of scattered light around the star is visible. Note, however, that this is not an example of the very bright halos that get entirely masked. An example of one of those is shown below. The black bars in each image represent 1$^{\prime}$.}
\label{fig:example_det}
\end{figure*}

\subsection{Step 3: Object Masking}
 In order to detect extremely faint candidate satellites we thoroughly mask the images in order to not be inundated with false positive detections. We mask both background galaxies and their associated diffuse light that can easily mimic diffuse LSB satellite galaxies. We do this following the method of \citet{johnny2}. In short, objects are detected in the image (note the image has already been masked for stars) at both a high and low detection threshold. The LSB detections are associated with an HSB detection if they overlap more than a certain fraction of their pixels with the HSB detection. The detected LSB pixels that are associated with an HSB detection get masked. These represent, for example, the extended envelopes of background galaxies, intracluster light from galaxy clusters, and/or scattered light around stars fainter than Gaia's completeness limit that were not masked in Step 1. The HSB detection threshold, the LSB detection threshold, and the overlap fraction parameters are generally around 15$\sigma$, $1\sigma$, and 0.1, respectively. 
 
 Due to the drastically differing imaging depths and level of star contamination across our sample fields, we were unable to use the same parameters for each host. The parameters that were required for a certain host to reduce false positives to reasonable levels would either let pass too many false positives in another region or significantly over-mask the region. In general the hosts with the shallowest data benefited from a lower LSB threshold (down to 0.5$\sigma$) and lower overlap fraction threshold (0.05). As discussed in \citet{johnny2}, there is a trade-off in determining the overlap fraction - a higher threshold lets in more background galaxies and their associated diffuse light, while a lower threshold makes it more likely that a real LSB satellite that overlaps with a bright point source (perhaps a foreground star or a physical nucleus of the LSB galaxy) will get masked. These parameters and the others described below are chosen after extensive experimentation for each region to try to minimize contaminants while not masking high-priority candidates that are found by eye in a subset of the surveyed area. 
 
 After this masking step, small faint sources (either MW foreground stars or small, background galaxies) remain in the image. These sources can blend together when we smooth the image and represent a significant contribution of false positives \citep[e.g.][]{sifon2018}. To deal with this, we detect and mask all sources that are $3-5\sigma$ above the background, depending on the depth of the data. For hosts with shallower data, this threshold is reduced to $3\sigma$, while for the hosts with the deepest data, this threshold is increased to $5\sigma$. We note that this will mask the central regions of many of the brighter LSB dwarfs but these dwarfs will have significant amounts of unmasked diffuse light surrounding the masked center that will still be detected. We note that \citet{bennet2017} employs a similar masking stage. It is possible that small and compact objects get completely masked during this stage, and this plays a role in setting the small-size completeness limit for each host shown in \S\ref{sec:completeness}.

 \subsection{Step 4: LSB Object Detection}
 With the images masked, we smooth the images with a Gaussian kernel with FWHM equal to the FWHM of the PSF for that region. For the fields with shallower data, a Gaussian kernel with twice the PSF FWHM is used instead to bring out the LSB galaxies. Then we run a final detection step detecting sources ${\sim}3\sigma$ above the background with size $>600$ pixels above the threshold (corresponding to a circle with radius ${\sim}2.5$\arcs). Again, the specific values depend on the host. Galaxies with deeper data have higher thresholds. Additionally, the closer hosts have a larger size cut due to the larger size that satellites in these systems would subtend on the sky. The detected objects are cut to those with average surface brightness within the effective radius fainter than 23.5 mag arcsec$^{-2}$. While this cut does not remove many sources, it does remove a few bright point-like contaminants that managed to evade masking in the previous steps. 
 
Figure \ref{fig:example_det} shows what detected objects look like on the masked and filtered images. Often the central regions of the dwarfs or peaks in the surface brightness fluctuations get masked but sufficient diffuse light remains to be detected.

 \subsection{Step 5: Visual Inspection}
 We require the objects to be detected independently in both filters and merge the list of HSB and LSB detected objects in each of the two filters. We consider a detection to be in both filters if the detections are within 4 times the average \texttt{FLUX\_RADIUS} (\texttt{SExtractor}'s approximation of the effective radius) of the two detections of each other. A large tolerance is required since often the diffuse envelope of a source is detected and a different side of the galaxy might get detected in the two filters.
 
 At this stage, the number of detected sources ranges from 62 to 1038 depending on the region. These correspond to roughly 100-200 candidates per square degree which is fairly similar to other semi-automated LSB galaxy detection algorithms \citep[e.g.][]{bennet2017, merritt2014}. Even with our aggressive masking, the majority of these detections are false positives that are roughly equal parts blends of small background galaxies and unmasked diffuse envelopes of bright background galaxies\footnote{Entire background galaxies that are detected in the HSB detection step contribute a smaller fraction of false positives.}. To eliminate these contaminants, we visually inspect all of the detections. As mentioned above, we also visually inspect the cutout areas around very bright stars that exhibit significant halos of scattered light.
 
 In performing the visual inspection, we focus on finding dwarfs that are morphologically diffuse and fairly regular.  Spiral arms in a small ($r_e<10$\arcs) galaxy and/or extremely bright cores likely indicate that a detection is background. Because our hosts are all within $D\lesssim10$ Mpc, bona fide dwarfs will exhibit noticeable mottling due to SBF in their surface brightness profiles in $r$ or $i$ because their stars are semi-resolved. We do not select on the presence of this SBF at this stage but wait until we perform a quantitative measure of the SBF to clean the detections. Still, very visible SBF means that we are not likely to reject real satellite galaxies at this stage, especially bright, blue ones, where the SBF will be strong and visible.
 
 In Appendix \ref{app:viz} we show some examples of rejected galaxies, along with the reason for their rejection. As we show below, we recover the vast majority of previously cataloged satellites in these regions. We take this as evidence that we are not performing too harsh of a selection in the visual inspection. There are two cases in which we visually reject a candidate on morphological grounds that had been previously cataloged as a probable satellite. In these cases,  our imaging data are significantly deeper than prior work and we are able to make a better judgement of the morphology of the candidate. Additionally, we note that our final sample of candidates are qualitatively quite similar in appearance to the final candidate samples of similar searches done by others \citep[e.g.][]{merritt2014, bennet2017}.

 As determined by the visual inspection, the false positive rate of our detection algorithm ranges from ${\sim}$9:1 to 31:1, among the galaxies that have a significant number of passable candidates (NGC 1023, NGC 4258, NGC 4565, NGC 4631, M51, M104). Again, this is comparable to other semi-automated LSB galaxy detection algorithms \citep[e.g.][]{merritt2014, bennet2017}.

\begin{figure*}
\includegraphics[width=\textwidth]{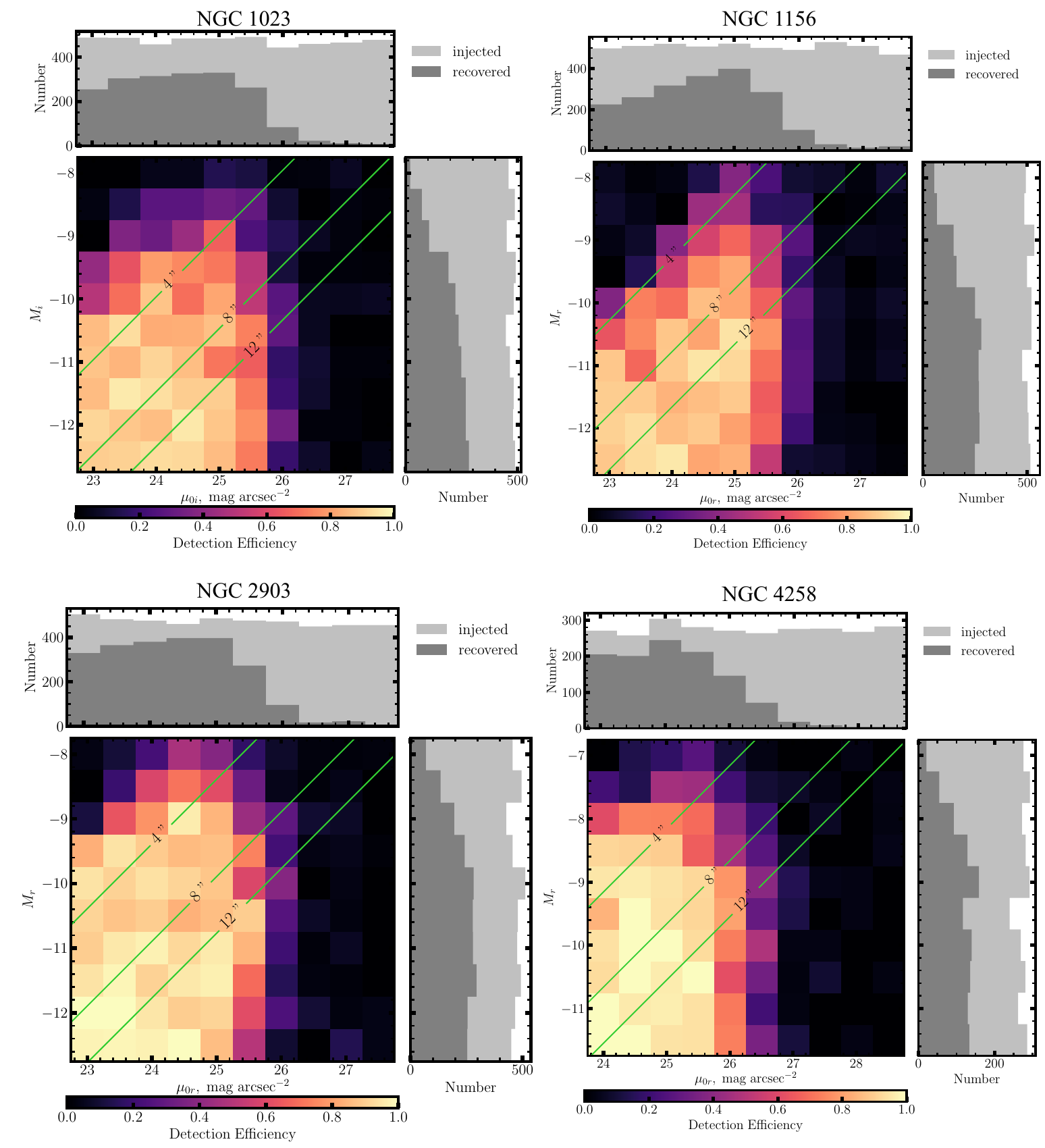}
\caption{Results of completeness tests for NGC 1023, NGC 1156, NGC 2903, and NGC 4258. Artificial galaxies are injected onto the chip-level images, and the recovery efficiency is measured as a function of total magnitude and central surface brightness. Also shown are the histograms of injected and recovered galaxies in magnitude and central surface brightness. Various contours of constant angular size are shown in the main panels.}
\label{fig:completeness1}
\end{figure*}

\begin{figure*}
\ContinuedFloat
\includegraphics[width=\textwidth]{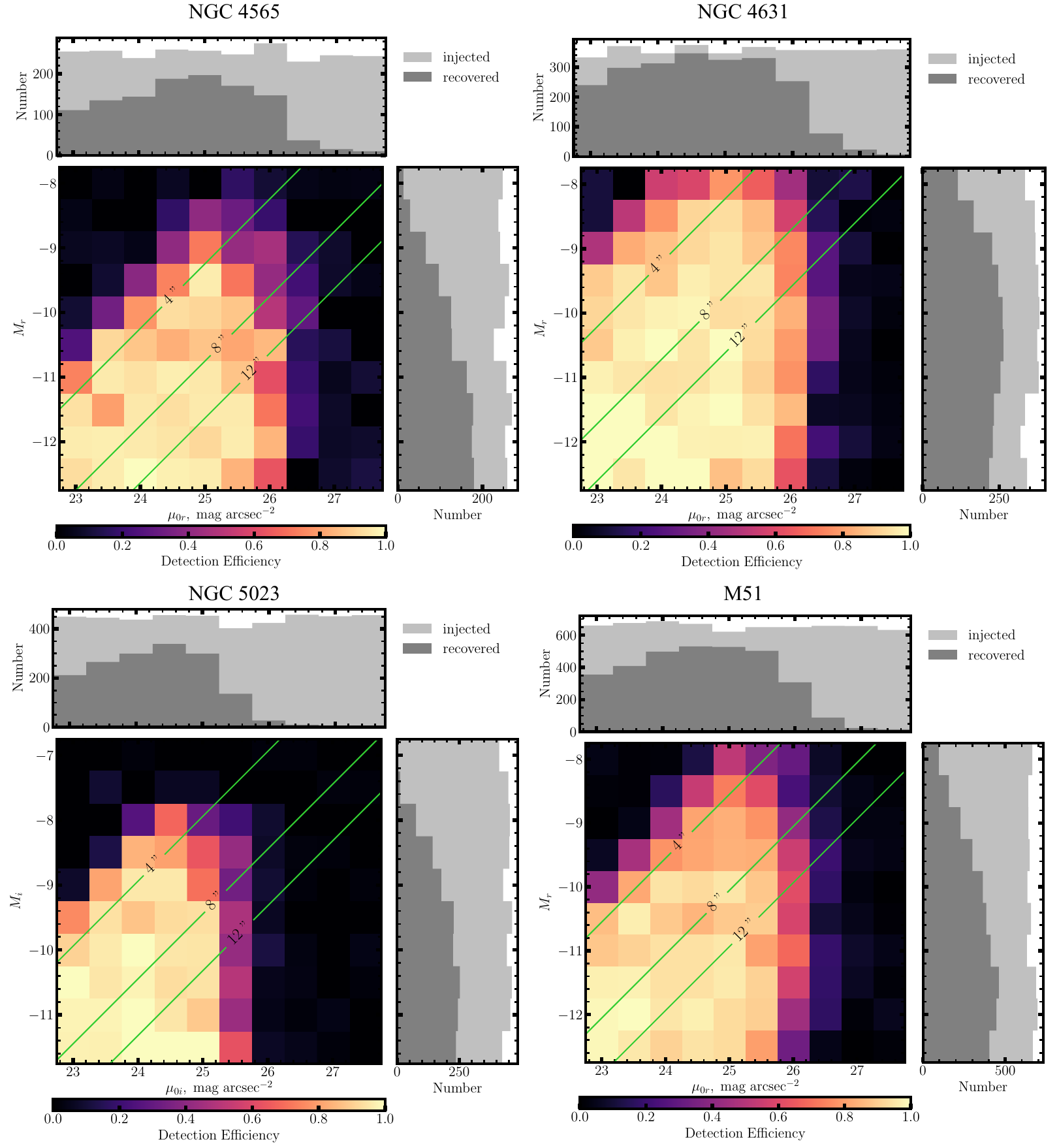}
\caption{The completeness tests of NGC 4565, NGC 4631, NGC 5023, and M51.}
\label{fig:completeness2}
\end{figure*}

\begin{figure*}
\ContinuedFloat
\includegraphics[width=\textwidth]{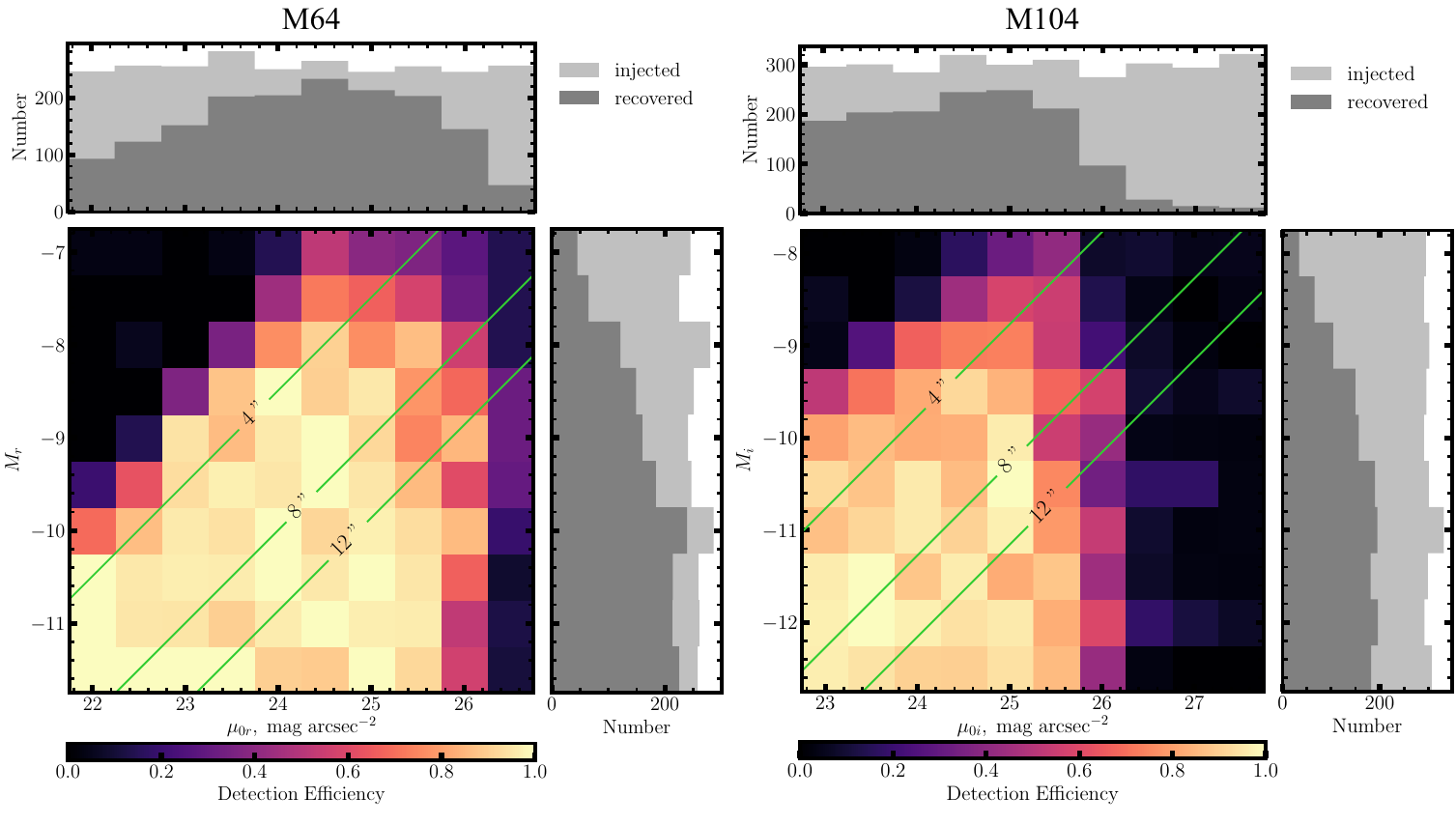}
\caption{The completeness tests of M64 and M104.}
\label{fig:completeness3}
\end{figure*}

\section{Completeness Tests}
\label{sec:completeness}
To compare this sample with theoretical predictions of satellite populations, it is crucial to understand and quantify the completeness of the detection algorithm. Additionally, as mentioned in \S\ref{sec:detection}, there are a number of free parameters in the detection algorithm that determine the sensitivity of the search, with values that are somewhat ad hoc. This is relatively unimportant as long as we test the detection algorithm using the same parameters as used when searching the hosts in our sample.

We perform completeness tests by injecting artificial galaxies of known magnitude and surface brightness and checking the recovery efficiency. We use $n=1$ S\'{e}rsics as models for the galaxies and vary the total $i$ or $r$ band magnitude and the central $i$ or $r$ band surface brightness. To determine the $g$ band galaxy properties, we assume either a $g-i$ color in the range $[0.4,0.8]$ mag or $g-r$ color in the range $[0.25,0.65]$ mag, which cover the range of colors of the detected candidates (see below). We inject the galaxies into the chip-level images, e.g., before background subtraction and co-addition, to incorparate any effect that the background subtraction has on detection. We inject ${\sim}40-60$ galaxies per 0.5$\times$0.5 mag wide bins in total magnitude and central SB. Galaxies are randomly placed in a circular area of the sky that completely encompasses the actual surveyed area (cf. Fig. \ref{fig:area1}). This is done to deal with the complicated and irregular survey areas. Once the galaxies are inserted onto the chips, we coadd the data as described above. Then the coadded tracts are run through the detection algorithm exactly as the real data are. Due to the very large number of fake galaxies, we do not perform the visual inspection step for the artificial galaxies but assume that any artificial galaxy that gets detected by the algorithm would pass through the visual inspection stage. We have verified that the lowest surface brightness artificial galaxies that the detection algorithm detects are still visible in the coadds and, hence, likely would be passed in the visual inspection.

The specific dithering patterns for each host often lead to certain areas being more deeply exposed than others. To prevent these areas from biasing the completeness tests, only galaxies that fall on a number of chips between the 10$^{th}$ and 90$^{th}$ percentile in the distribution of coaddition stack depth are used to calculate the completeness. We additionally exclude galaxies that land in the large scattered light halos around the very bright stars. These large areas are visually searched in the real data and we address the completeness for these areas below. Note that galaxies that fall in the regular star-masked areas (the areas that do not get visually inspected) are counted so that we can quantify the area lost due to star-masking. The whole injection, coaddition, and detection process is repeated 5-10 times per host to build up ample statistics. 

Figure \ref{fig:completeness1} shows the results for each galaxy. The detection efficiency throughout the input grid of total magnitude and central surface brightness is shown along with marginal distributions for the magnitude and SB. The results for each of the ten hosts are qualitatively quite similar. Completeness is close to 1 for artificial galaxies with the highest total magnitude and central surface brightness. There is a steep drop-off in completeness at a certain faint central surface brightness that depends on the host and the filter ($r$ or $i$). In the $r$ band, this cutoff ranges from ${\sim}25.5$ mag arcsec$^{-2}$ for NGC 1156 to ${\sim}26.5$ mag arcsec$^{-2}$ for NGC 4258 and NGC 4631. In the $i$ band, this cutoff ranges from ${\sim}25$ mag arcsec$^{-2}$ for NGC 5023 to ${\sim}26$ mag arcsec$^{-2}$ for M104. There is also a cutoff to the completeness for low magnitudes at constant central surface brightness. This dropoff closely follows the contours in size, which are also shown in the plots for certain values of the angular size of the galaxies. Generally the cutoff in size is around $r_e{\sim}3$\arcs~which is close to the size cut used in the detection step described in \S\ref{sec:detection} (note that the size cut used varies between the hosts). 

Most galaxies have peak completeness $\gtrsim$90\%, with the exception of NGC 1023 and NGC 1156 which are closer to ${\sim}80$\%. For NGC 1023, this is due to the large number of stars and the corresponding area lost to the star-masking. For NGC 1156, this is due to the large amount of cirrus in the field which required us to use a higher detection threshold or otherwise be inundated with false positives.

\begin{figure*}
\includegraphics[width=\textwidth]{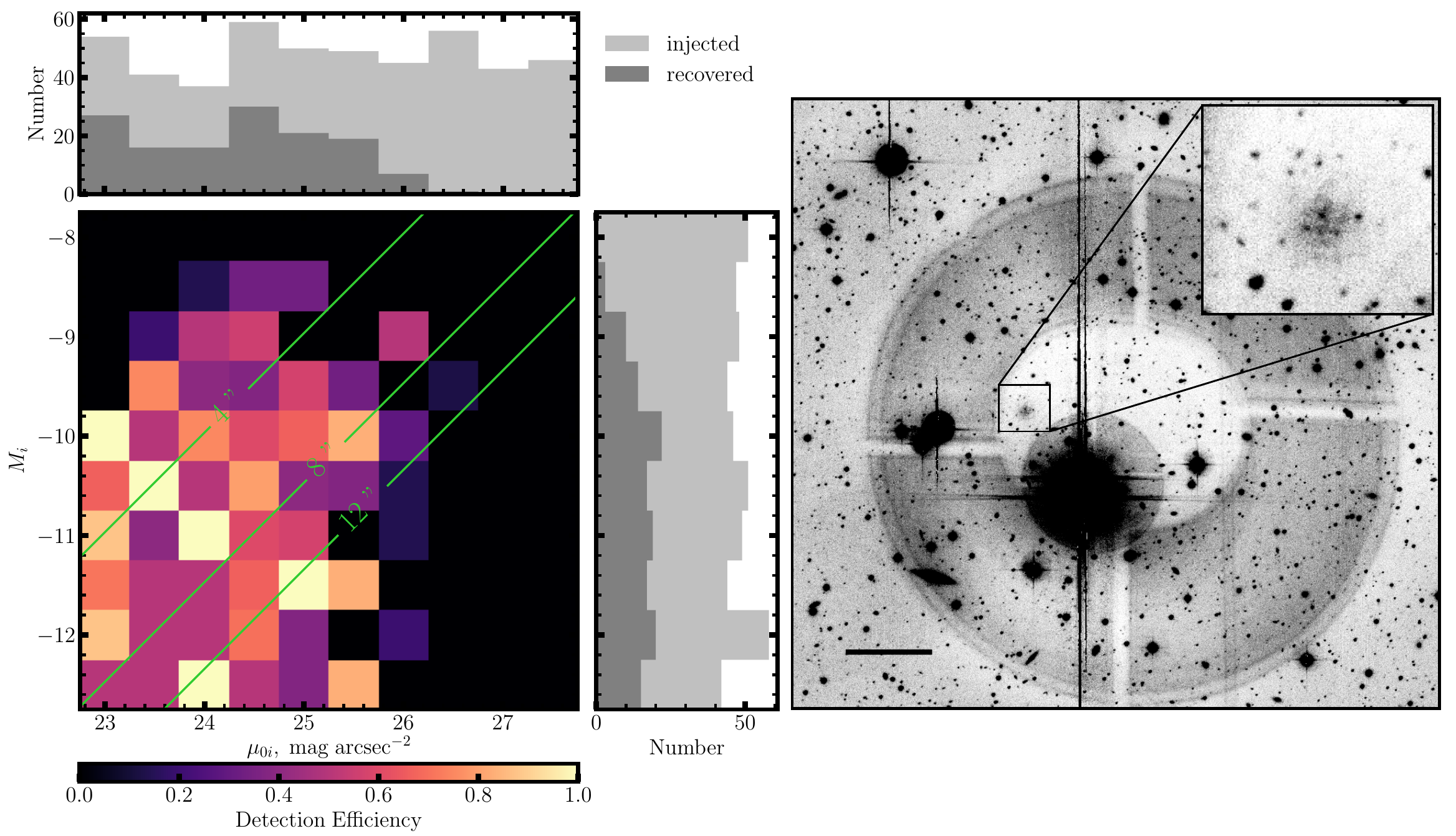}
\caption{The completeness tests for the cutout patches in the NGC 1023 field that get visually searched. Shown on the right is an example of a scattered light halo in the NGC 1023 field that gets masked. The inlay shows an example of a high priority satellite candidate that is found from visually searching cutouts of the masked halos. The black bar represents 1$^\prime$.}
\label{fig:completeness_viz}
\end{figure*}

As mentioned above, we visually search through the large masked (${\sim}8^{\prime}$) areas around the brightest, $\lesssim9^{th}$ magnitude stars. While these areas generally constitute $\lesssim$5\% of the total surveyed area around each host, it is still important to quantify the completeness of the visual searching. The exception is NGC 1023, where the masked areas constitute ${\sim}$10\% of the area. To check the efficiency of the secondary visual searching, on each iteration of the completeness checks for the automated algorithm described above, we visually search through the star mask cutouts in the NGC 1023 region. We show the recovery efficiency for this search in Figure \ref{fig:completeness_viz}. The efficiency is qualitatively similar to that shown for the detection algorithm in Figure \ref{fig:completeness1} with similar surface brightness and size bounds. The peak efficiency is significantly lower at ${\sim}$50\%. This is unavoidable due to the amount of scattered light in these halos. However, this shows that many galaxies are still detectable by eye in these cutouts. Considering how little area these cutouts constitute out of the entire surveyed areas and the fact that recovery is still ${\sim}$50\%, we consider that Fig. \ref{fig:completeness1} accurately represent the overall recovery efficiency of LSB galaxies in the different regions.

\section{Catalogs of Detected Satellite Candidates}
\label{sec:catalogs}

\subsection{Overview}
We present the catalogs of dwarfs detected in the host fields that passed visual inspection in Tables \ref{tab:ngc1023}-\ref{tab:m104}. In total, 153 galaxies pass visual inspection. The number of dwarfs in each region that pass visual inspection varies widely between the different hosts from $>30$ around NGC 4258 and NGC 1023 to just one around M64. Table \ref{tab:overview} gives some overview numbers for the candidate sample in each region.

\begin{deluxetable*}{ccc}
\tablecaption{Overview of Detected Candidates\label{tab:overview}}

\tablehead{\colhead{Name} & \colhead{\# of Candidates} & \colhead{Sq. Deg. Coverage}} 
\startdata
 NGC 1023 & 31 & 3.7   \\
 NGC 1156 &  3 & 0.91       \\
 NGC 2903 &  4  & 1.7       \\
 NGC 4258 &  33 & 4.3        \\
 NGC 4565 &  21 & 2.5       \\
 NGC 4631 & 16  & 3.3        \\
 NGC 5023 &  2 &  1.1      \\
 M51 &  15 &   3.2          \\
 M64 &  1 &  1.8        \\
 M104 &   27 &  1.8        \\
\enddata
\end{deluxetable*}

The photometric properties of the galaxies come from our measurements with the CFHT data, unless otherwise indicated. Galaxies with TRGB distances available are indicated, as are galaxies with available redshifts. Galaxies with NGC, IC, or UGC catalog names are referred to by those names. Other dwarfs are named by their RA/DEC. Where available, previous catalog names are given along with references. In total, 93 appear to be new detections\footnote{By ``new'' we mean simply that the galaxy has not previously been recognized as a satellite or candidate satellite. We also do not include any candidate that has an archival redshift in this category.}. We compare in more detail with previous searches below, but we mention here that we detect everything from previous searches that falls within our footprints with the exception of a small dwarf around NGC 4631 that is projected on top of the disk of NGC 4631. There are no previous detections that are too low surface brightness for us to detect with the MegaCam data.

\begin{figure*}
\includegraphics[width=\textwidth]{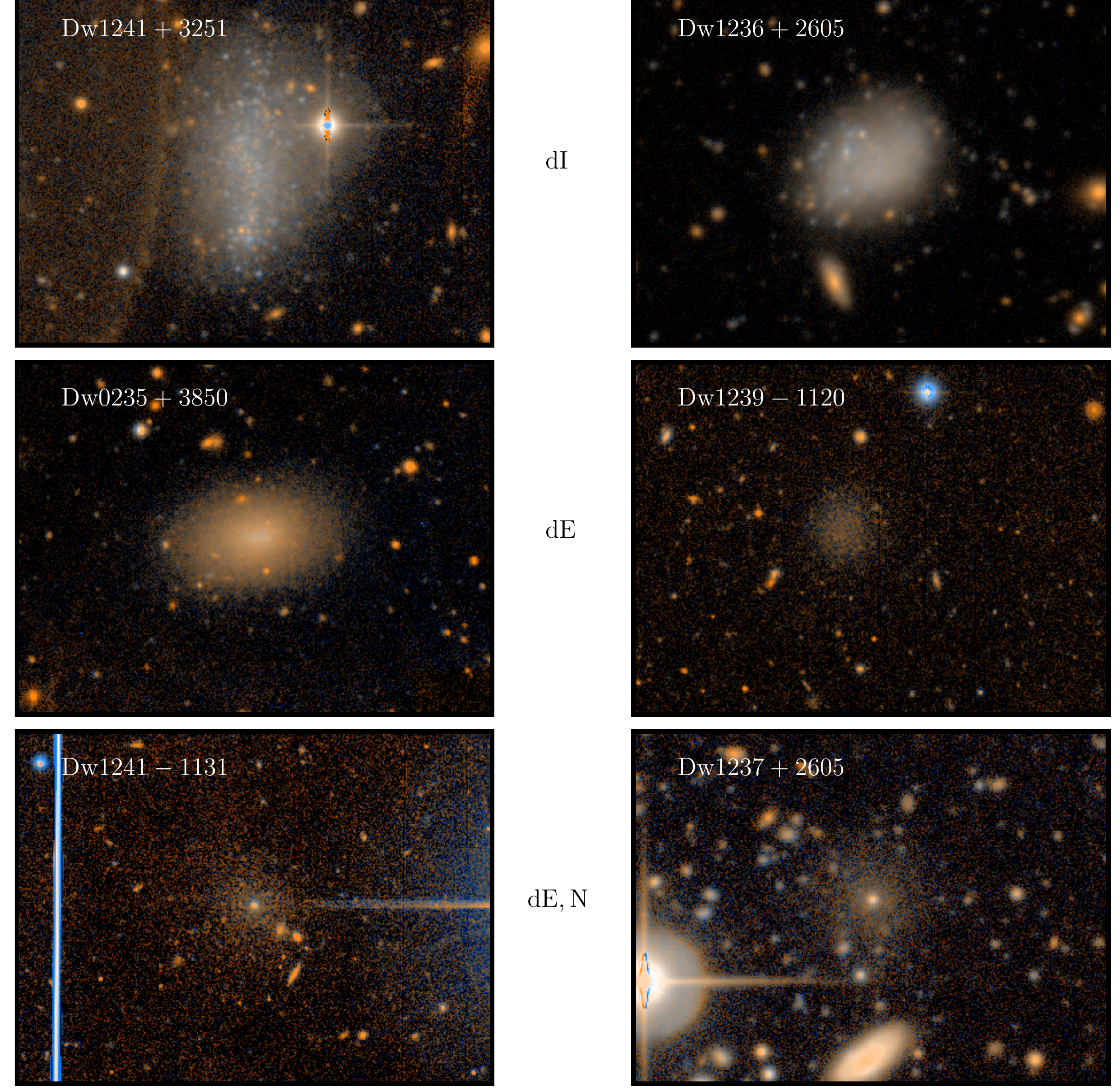}
\caption{Color composite images for six select dwarfs out of the entire detected sample. Each image uses either the $r$ or $i$ band image for the red channel, the $g$ band image for the blue channel, and the average of the two for the green channel. Each panel is $100$\arcs~ wide. The top two dwarfs are classified as dI, the middle two as dE, and the bottom two as dE,N. Several of the dwarfs exhibit very noticeable SBF. }
\label{fig:select_dws}
\end{figure*}

Tables \ref{tab:ngc1023}-\ref{tab:m104} list a rough galaxy type for each detected candidate. The types are estimated from the morphology shown in the CCD images. Generally the type is chosen from ``dI", ``dE", or ``dI/dE" for ambiguous cases. A few of the larger detections are classified as ``Im" or ``Sdm". Additionally, an ``N" is included if the dwarf appears to be nucleated. Here, we loosely define `nucleated' as having a bright point source within 1-2\arcs~of the center of a dwarf. Figure \ref{fig:select_dws} shows color images of six example dwarfs, two each of the types dI, dE, dE,N.

\input{ngc1023.tex}

\input{ngc1156.tex}

\input{ngc2903.tex}

\input{ngc4258.tex}

\input{ngc4565.tex}

\input{ngc4631.tex}

\input{ngc5023.tex}

\input{m51.tex}

\input{m64.tex}

\input{m104.tex}

\subsection{Photometry}

To derive the photometric quantities listed in Tables \ref{tab:ngc1023}-\ref{tab:m104}, we re-reduce cutouts around each of the detected dwarfs. The goal of the re-reduction is to perform a better sky subtraction that gives more accurate photometry and is more suitable for the SBF analysis that will be presented in a future paper. In this re-reduction, we find all the chips that either cover the galaxy or are very close to it and photometrically and astrometrically calibrate these chips. As mentioned in \citet{sbf_calib}, the SBF measurement of LSB galaxies is very sensitive to the sky subtraction, and sky subtraction that is too aggressive will bias the SBF measurements to larger fluctuation levels. \citet{sbf_calib} used a sky subtraction algorithm that was loosely based on that of \texttt{Elixir-LSB} of \citet{ngvs} and showed with simulations that it accurately recovers colors and SBF magnitudes. However, here we opt for a simpler process. For the sky subtraction in this re-reduction we mask a circle around each detected dwarf (generally 200-300 pixels in radius) and then use \texttt{SExtractor} to estimate the sky background with a 512$\times$512 pixel mesh. In image simulations, we find that the use of such a large mesh and the masking of the dwarfs prevents the partial subtraction of the dwarfs that plagues more aggressive, local sky subtractions and that we can recover SBF magnitudes, colors, and integrated magnitudes at least as well as the algorithm used in \citet{sbf_calib}.

Once cutouts for each detected dwarf are re-reduced with the better sky subtraction, S\'{e}rsic profiles are fit to the galaxies using \textsc{imfit} \citep{imfit}. Due to $g$ band generally being deeper than either $r$ or $i$, the fits are done in the $g$ band with all the S\'{e}rsic parameters left free and then the redder bands are fit with only the amplitude of the profile free to vary. For dwarfs that appear to be nucleated, the nucleus is masked in the fitting. The photometric quantities reported for each dwarf generally come from these fits unless otherwise indicated.  The uncertainties in the magnitudes, colors, sizes, and surface brightness come from fits to artificial galaxies that are placed in the data in fields nearby to the real galaxies. The real galaxies are fit with S\'{e}rsic profiles and then these fits are used to generate artificial galaxies that are then inserted into the data at the CCD level. This is before sky subtraction and so the uncertainties will accurately reflect the effect of the sky subtraction process that we use. The artificial galaxies are then fit with S\'{e}rsic profiles and the spread in the recovered values is used as an estimate of the uncertainty in the quantity for the real galaxy. We find that colors are generally recovered with precision $\lesssim0.1$ mag. The magnitudes in $g$, $r$, and $i$ are generally recovered within ${\sim}0.2$ mag, which is worse than the color due to covariance in the magnitude in the different filters. We assume a minimum photometric error of 0.01 mag which we estimate is roughly the accuracy of the photometric calibration.

\subsection{Discussion of Individual Hosts}
In this section, we discuss each host, giving relevant details about each host and outlining any previous work on characterizing the satellite systems. Where there are prior searches, we compare our sample of candidates with the previous samples.

\subsubsection{NGC 1023}
The NGC 1023 group is a spiral rich group at ${\sim}$10.4 Mpc  (median distance in NED) that is dominated by the lenticular galaxy, NGC 1023. Its low galactic latitude $|b|{\sim}20$ makes it a less ideal target due to a profusion of bright MW stars (see Section \ref{sec:detection} for a discussion of star-masks), but the great depth of the imaging data partially compensates for it. \citet{trentham2009} estimate its dynamical mass as $6.4\times10^{12}$ \msun~ from the kinematics of the bright members that have spectroscopic observations. \citet{trentham2009} cataloged the dwarf members of this group using very wide-field (60 sq. degrees) $r$ band CFHT/MegaCam imaging that had exposure times of 10 minutes per field, estimating a completeness limit of $M_r{\sim}-10$. The search was visual and the association of dwarfs was determined by morphology alone for most cases. In our search, we do not use the same $r$ band data and opt to use much the deeper $g$ and $i$ band imaging of \citet{ngc1023_ucd}; this only covers the ${\sim}4$ sq. degrees around NGC 1023, but has $>3$ hours of exposure time per field in $i$ band. 

Of the ${\sim}60$ detected candidates of \citet{trentham2009}, 14 fall completely within our footprint. All of these are detected. We note, however, that their candidate [TT09]29 did not pass our visual inspection. Our imaging data is significantly deeper than that used by them and we detect surrounding LSB features that show sharp edges and twists that are not consistent with being a low mass galaxy at $D=10$ Mpc. This galaxy is shown in Appendix \ref{app:viz}. \citet{trentham2009} only considered this galaxy as a ``plausible'' member based on its morphology (as opposed to ``likely'' or ``possible''). Additionally, it shows no visible SBF which is easily seen in many of the other candidate satellites in this group. This means that 20 out of our 33 candidates are new detections. Many of the new detections are small and/or very LSB and likely not visible in the shallower data of \citet{trentham2009}. However, we do note a couple bright $M_g{\sim}-12$ candidates with strong SBF (and, hence, very likely real satellites) that were not cataloged in \citet{trentham2009}.

\subsubsection{NGC 1156}
NGC 1156 is an isolated dwarf irregular galaxy at $D=7.6$ Mpc \citep{kim2012}. Its closest neighbors appear to be more than 10$^\circ$ away \citep{kara1996} making it one of the most isolated galaxies in the Local Volume. Additionally, its current high rate of star formation does not appear to be triggered by any interaction with companions \citep{hunter2004}. \citet{minchin2010} find a candidate dwarf in H\textsc{I} with Arecibo. The dwarf is at a ${\sim}80$ kpc projected distance from NGC 1156, unfortunately placing it barely outside of our survey footprint. From the relative velocity of the two galaxies and their separation, \citet{minchin2010} estimate a lower limit of ${\sim}1.1\times10^{11}$ \msun~for the dynamical mass of the NGC 1156 system. \citet{karachentsev2015} discovered two more LSB candidate satellite systems in the vicinity of NGC 1156 from deep small-telescope imaging around NGC 1156, and both of these targets are in our footprint. Both are easily recovered along with a third small, irregular dwarf candidate.

\subsubsection{NGC 2903}
NGC 2903 is a barred spiral at $D=8.0$ Mpc (mean NED distance). It is in the isolated galaxy catalog of \citet{karachentseva1973}, being located at the edge of the Gemini-Leo Void. It appears to be quite similar to the MW in both SFR \citep{irwin2009} and H\textsc{I} rotation curve \citep{begeman1991}. \citet{irwin2009} list four candidate companions of NGC 2903, including one discovered by them with Arecibo H\textsc{I} observations. Two of these (including the H\textsc{I}-discovered dwarf) are in our footprint while the other two are substantially farther away ($>$240 kpc projected) from NGC 2903. \citet{karachentsev2015} searched for more LSB satellites using deep, small-telescope data sets but did not uncover any other promising candidates. We recover the two previously cataloged satellites along with two additional small, LSB satellite candidates.

\subsubsection{NGC 4258}
NGC 4258 (M106) is a barred spiral at $D=7.2\pm0.2$ Mpc \citep{m106_maser}. It has a peak rotation speed of 208 km/s \citep{m106_vcirc}, slightly less than that of the MW. It is the dominant galaxy of the CVn \textsc{II} group. Several dwarf companions are already known. \citet{m106_vcirc} confirmed NGC 4248 as a bright companion via an H\textsc{I} redshift and \citet{jerjen_field} confirmed UGC 7356 to be at the distance of NGC 4258 via an SBF measurement (later confirmed with TRGB). \citet{kim2011} performed a wide-field search with CFHT/MegaCam and visually found 16 candidate companions. Two of the sample of \citet{kim2011} were shown by \citet{cohen2018} to be background contaminants while one was confirmed to be at the distance of NGC 4258. \citet{cohen2018} further confirm two more nearby LSB galaxies to be background. One was first noted by \citet{binggeli1990} and the other found by \citet{karachentsev2015}\footnote{Both of these candidates fell in chip gaps in the original survey of \citet{kim2011} and were not noted there.}. 

Comparing our sample to that of \citet{kim2011}, we detect all of their 16 main sample of candidates. We did not include three of these (S2, S3, and S15) in our final sample because these have redshifts of ${\sim}800-900$ km/s which are substantially larger than that of NGC 4258 (460 km/s) and are very large, massive HSB spirals. These galaxies are almost certainly background and primary galaxies in their own right. \citet{kim2011} list another 5 candidates as lower priority ``possible'' satellites, and, of these five, P1 and P5 are in our final sample. Two of the other three were smaller than our detection size limit while the third did not pass visual inspection due to its twisted, irregular shape. We recover one of the two candidates of \citet{cohen2018} that fell in chip gaps for \citet{kim2011} since we supplement with additional MegaCam data to fill in some of the chip gaps. The fifth object (NGC 4258-DF5) that \citet{cohen2018} investigate still falls on a chip gap in our survey footprint. We also recover all objects catalogued by \citet{kkh07} that are in our footprint.

\citet{spencer2014} used SDSS targeting and spectroscopic follow-up to confirm satellite membership. As mentioned by \citet{karachentsev2015}, however, this approach is complicated by the presence of at least one background group along the line of sight that is at a similar recessional velocity as NGC 4258. Out of the 8 candidates that they consider to be spectroscopically confirmed satellites, 7 are in our footprint. 4 of these are dwarfs in the sample of \citet{kim2011} and, hence, are also in our final sample. Two of the remaining three are in our final sample while the third (SDSS J121551.55+473016.8 in their catalog) did not pass the visual inspection. It has the appearance of a fairly HSB, edge-on disk with no visible SBF in the outskirts and is very likely background, perhaps belonging to a background group along the line of sight. This candidate is shown in Appendix \ref{app:viz}.

Most of our new detections are fairly small and compact and it is possible that the previous visual searches simply missed them. It is likely that most of them will turn out to be in the background. We do note that a few of them have the LSB, diffuse morphology expected of a satellite and are promising candidates. In particular, we highlight the very LSB dw1218+4623 as an auspicious candidate.

\subsubsection{NGC 4565}
NGC 4565 is an edge-on spiral at $11.9^{+0.3}_{-0.2}$ Mpc \citep{ngc4565_dist}. It is slightly more massive than the MW with a peak rotation speed of 244 km/s. \citet{kormendy2019} point out that NGC 4565 is a prime structural analog to the MW due to the very similar properties of its pseudobulge\footnote{We note that a pseudobulge is simply a bulge that forms from secular processes as opposed to the merger origin of a classical bulge \citep{kormendy2004}.} to that of the MW. This makes it an interesting target to explore how satellite system properties correlate with the bulge of the host spiral \citep[e.g.][]{lopez_bulge, javanmardi_bulge}. Its satellite system is previously unexplored. 

We detect 21 LSB candidate satellites around NGC 4565. A few of these have SDSS spectra but most are small and LSB. The spatial arrangement shown in Figure \ref{fig:area1} is lopsided with many candidates to the north of NGC 4565. We suspect this is due to the background group centered on NGC 4555 (z${\sim}0.02$) that is rich in LSB galaxies. This would fit with many of the candidates being particularly small.

\subsubsection{NGC 4631}
NGC 4631 is an edge on spiral at $D=7.4\pm0.2$ Mpc \citep{ngc4565_dist} currently undergoing an interaction with NGC 4656 and perhaps NGC 4627, revealed through significant disturbances present in H\textsc{I} \citep{rand1994}. The tidal material near NGC 4656 might host a tidal dwarf galaxy \citep{ngc4631_tdg}. There is a large tidal tail that aligns with the direction from NGC 4631 to NGC 4656 \citep{dmd2015} but is not due to the interaction between these large spirals or associated with the H\textsc{I}. Instead, it is likely from the accretion of a smaller satellite. A number of LSB galaxies have been detected in the region from amateur-telescope imaging \citep{dmd2015, javanmardi_m101, karachentsev2015}. \citet{tanaka2017} used extremely deep Subaru/HSC imaging to study the stream and cataloged a total of 11 dwarf satellite candidates (including the previously discovered dwarfs). The CFHT/MegaCam imaging we use is not as deep as the HSC imaging, but it is slightly wider field.

We recover nearly all of the satellite candidates cataloged by \citet{tanaka2017}. This includes DGSAT-1 and DGSAT-2 from \citet{javanmardi_m101}. We concur with \citet{tanaka2017} that DGSAT-3 from \citet{javanmardi_m101} was a blend of some foreground stars and background galaxies, and that no real LSB dwarf exists there. The one candidate cataloged by \citet{tanaka2017} that we do not recover is their HSC-4 which is a small grouping of young, massive stars projected on top of the disk of NGC 4631. This dwarf was first noticed in the \emph{HST} imaging of \citet{seth2005}. This dwarf appears to consist almost entirely of bright, blue stars without any diffuse background component. This fact compounded with the dwarf's projected placement in front of the disk of NGC 4631 made it undetectable by our detection algorithm, which picks up on diffuse light. There are several candidates in our final sample that are not in that of \citet{tanaka2017}. Three of these (dw1242+3231, NGC 4627, and dw1240+3247) are in their footprint. dw1242+3231 is a compact system that is also projected on top of the disk of NGC 4631. Unlike HSC-4, it has ample diffuse light that makes it detectable by our algorithm. NGC 4627 is the very bright companion directly north of the disk of NGC 4631. Since it is being tidally distorted by NGC 4631, its association as a satellite is very certain. dw1240+3247 is the progenitor of the tidal stream to the northwest of NGC 4631. \citet{tanaka2017} map this stream in great detail but do not include the progenitor in their satellite sample. Finally, we have three satellite candidates that are outside of the HSC imaging footprint. Two of these (dw1239+3230 and dw1239+3251) have diffuse morphologies and visible SBF and we believe they are likely satellites. The third, UGCA 292, is fully resolved and has a literature TRGB distance that puts it significantly in the foreground. 

We note that the detected candidate satellites around NGC 4631 are very centrally clustered compared to the other hosts.

\subsubsection{NGC 5023}
NGC 5023 is an edge-on dwarf galaxy at $D=6.5\pm0.2$ Mpc \citep{ngc4565_dist}. It has a maximum circular rotation speed of 89 km/s \citep{ngc5023_rot}. It is near in projection to M51 and M63 but significantly (2 Mpc) closer and is not associated with either massive primary \citep{t15}. The LV catalog of \citet{karachentsev} lists M101 as the `Primary Disturber' of NGC 5023. While NGC 5023 is at a very similar distance as M101, they are $\gtrsim500$ kpc apart in projection. NGC 5023 appears to be a fairly isolated LMC-mass dwarf. We detect two small dwarf candidates in the area.

\subsubsection{M51}
M51 (NGC 5194) is a face-on spiral at $D=8.6\pm0.1$ Mpc \citep{m51_dist}. With a flat rotation curve speed of ${\sim}220$ km/s \citep{m51_vcirc1,m51_vcirc2}, it is roughly similar in mass to the MW. In addition to the obvious interaction with NGC 5195, M51 shows extensive LSB tidal debris which might hint at other recent accretion events \citep{m51_tidal}. \citet{t15} argue that M51 and M63 form a gravitationally bound group akin to the MW and M31 with a projected separation of ${\sim}$900 kpc. \citet{t15} also note that most dwarfs known in the vicinity of M51/M63 are foreground, leaving the satellite system of M51 relatively unexplored \citep[see also][]{muller101}. 

We catalog 15 candidates in this region. This includes NGC 5229 which is the only possible satellite that the Local Volume Catalog of \citet{karachentsev} lists that is in our footprint. Interestingly, our sample also includes one of the seven LSB galaxies from \citet{dalcanton1997}. Many of the candidates appear concentrated to the south of M51, which we believe is due to contamination from a background group surrounding NGC 5198.

\subsubsection{M64}
M64 (NGC 4826) is a spiral at $D=5.3$ Mpc \citep{m64_dist}. It is relatively isolated with no known massive neighbors \citep{turner1976, brunker2019}. With a peak rotation circular velocity of ${\sim}155$ km/s \citep{rubin1994}, it is distinctly less massive than the MW. While it appears to be currently fairly isolated, the presence of a counter-rotating outer gas disk is indicative of a previous merging event \citep{braun1994}. While the area coverage of our search is not ideal, the fact that its satellite system is currently unexplored makes M64 a worthwhile addition to our sample.

We detect only a single dwarf candidate in this region. While the area surveyed is quite small, the completeness (cf. Figure \ref{fig:completeness1}) is quite good. We will explore in later papers whether this dearth of satellites is expected given the survey area.

\subsubsection{M104}
M104 (NGC 4594; Sombrero Galaxy) is a giant elliptical \citep{m104_elliptical} at $D=9.6\pm0.1$ Mpc \citep{m104_dist}. It is along the line of sight to a major southern spur of the Virgo cluster but is significantly in the foreground. With an estimated asymptotic circular rotation speed of 380 km/s \citep{m104_vcirc} and stellar mass of ${\sim}1-2\times10^{11}$ \msun~\citep{m104_halo, m104_stars}, M104 is significantly more massive than the MW. \citet{m104_halo} estimate its dynamical mass as ${\sim}3\times10^{13}$ \msun~from the kinematics of 6 nearby dwarfs. Most of these dwarfs are quite distant from M104, with projected separations $>$400 kpc. One UCD is currently known in the system, SUCD1 \citep{m104_ucd}. \citet{javanmardi_m101} studied the inner satellite system with deep amateur imaging and reported three LSB satellite candidates which we recover. With the MegaCam imaging, we reveal a very large population of LSB satellite candidates. Many of these have easily visible SBF (cf. Figure \ref{fig:select_dws}) and are likely real satellites.

\begin{figure*}
\includegraphics[width=\textwidth]{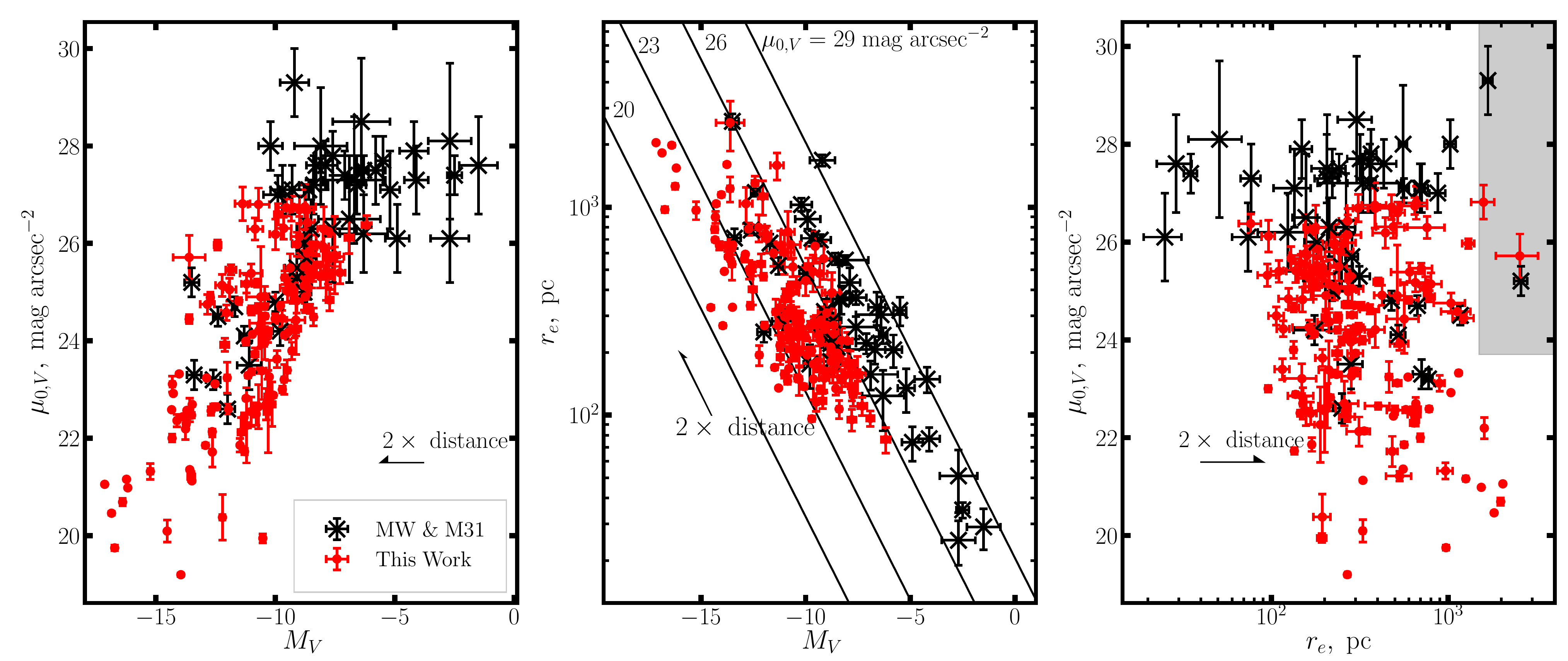}
\caption{The structural parameters of the sample of detected candidates in all surveyed hosts compared with the MW and M31 satellite sample of \citet{mcconnachie2012}. The dwarfs assume the distance of the host that they were found around when calculating absolute luminosity and size. The contours of central surface brightness in the middle plot are calculated assuming an $n=1$ S\'{e}rsic profile. The arrows show the effect of moving a dwarf to twice the distance. The shaded region in the right plot is  $\mu_{0,g}>24$ mag arcsec$^{-2}$ and $r_e>1.5$ kpc, which is the usual definition of an ultra-diffuse galaxy.}
\label{fig:lg_struct}
\end{figure*}

\section{Discussion}
\label{sec:discussion}
It is likely that many (or even most) of the detected satellite candidates will turn out to be background contaminants. Indeed, a few candidates are already known to be background from \citet{cohen2018} or from available redshifts. We therefore wait to do a detailed analysis of the luminosity functions of the satellite systems and comparison to models until distance information is available for more of the candidates. SBF distances to the bright and large candidates will be presented in a forthcoming paper. We present an overview of the properties of the candidates in this section. In what follows, we assume the distance of the host for each dwarf to derive absolute physical quantities.

\subsection{Size, Luminosity, and Surface Brightness}
Figure \ref{fig:lg_struct} shows the structural parameters of the detected candidates compared with the satellite systems of the MW and M31 as tabulated by \citet{mcconnachie2012}. In making this plot, the distance to the host is assumed for each dwarf to derive the absolute luminosity and size. To calculate photometry in the $V$-band, filter transforms derived for SDSS are used\footnote{Found here: \url{http://www.sdss3.org/dr8/algorithms/sdssUBVRITransform.php}.}. Three different projections of the luminosity - size - surface brightness relation are plotted. In general there is fair qualitative agreement between the sample of candidates presented here and the known satellite systems of the MW and M31. Our completeness limit at $\mu_0\lesssim26.5$ mag arcsec$^{-2}$ is seen clearly in all three plots. This agrees well with what we find with the artificial galaxy simulations presented in \S\ref{sec:completeness}. Also shown in each plot is the effect that moving a dwarf to twice the distance would have on its location in the plot. From these plots, it is apparent that many of the candidates are background. At fixed absolute magnitude, there appears to be more candidates with small size (equivalently high surface brightness) than there are in the MW and M31 satellites. If many of these candidates were actually background, the agreement would be much better. 

Also shown in the right panel of Figure \ref{fig:lg_struct} is the region of parameter space occupied by so-called ultra-diffuse galaxies \citep{pvd2015}: $\mu_{0,g}>24$ mag arcsec$^{-2}$ and $r_e>1.5$ kpc. Only two detected candidates meet these criteria, dw1240+3247 in the NGC 4631 field and dw1235+2606 in the NGC 4565 region. dw1240+3247 is being tidally disrupted by NGC 4631 as evidenced by a tidal tail extending from it and pointing towards NGC 4631 \citep{dmd2015, tanaka2017}. dw1235+2606 also appears to be tidal due to its arc-like shape bending around the outer disk of NGC 4565. The absence of (non-tidal) ultra-diffuse galaxies around these hosts is in line with expectations from published relations between the number of ultra-diffuse galaxies and halo mass \citep[e.g.][]{vanderburg2017}. For halos with mass $<10^{13}$\msun~, one or less ultra-diffuse galaxies is expected. With the exception of M104, all of our primaries have a halo mass less than this.

\begin{figure}
\includegraphics[width=0.49\textwidth]{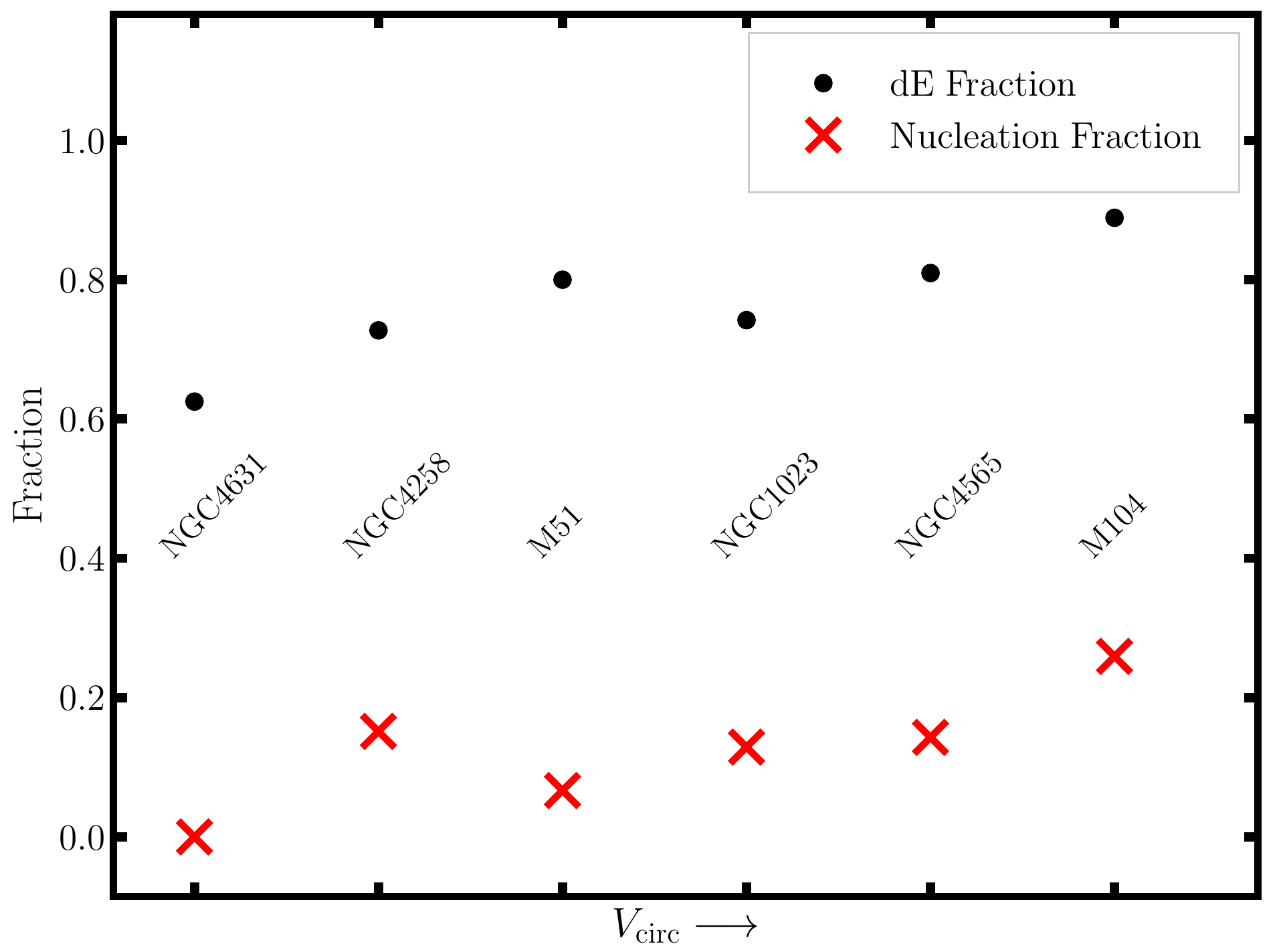}
\caption{The fraction of all detected candidates that are of type ``dE'' and the fraction of type ``dE'' candidates that have a noticeable nuclear star cluster. The hosts are in order of increasing peak circular rotation speed.}
\label{fig:type}
\end{figure}

\subsection{Nucleation and Type}
From the images we estimate the galaxy type and whether it has a nuclear star cluster (NSC) or not. While these visual classifications are fairly subjective, the same criteria are used for each host, making the comparison between hosts somewhat meaningful. In Figure \ref{fig:type}, we plot the fraction of candidates that are classified as ``dE'' and the fraction of dE candidates that are seen to have a NSC. Only the six hosts in our sample that have significant numbers of detected candidates (NGC 1023, NGC 4258, NGC 4565, NGC 4631, M51, and M104) are shown. A majority of candidates appear morphologically like quenched spheroid systems, and there does not appear to be much difference between hosts in the dE fraction. This high fraction of spheroid satellites agrees well with the satellite system of the MW where all known satellites except the LMC and SMC are spheroids. A similar majority of spheroid galaxies are found in the satellite systems of Cen A \citep{crnojevic2019} and M81 \citep{chiboucas2013}. This is somewhat in tension with the results of the SAGA Survey \citep{geha2017}, which has surveyed the satellite systems of 8 MW-analog hosts in the distance range 20-40 Mpc. While it is difficult to gauge morphology from the shallow SDSS imaging that SAGA uses for targeting, especially at the distance of the hosts surveyed, a majority of their confirmed satellites would be classified as irregular using the criteria used here. A more quantitative comparison with SAGA is possible with the colors of the detected candidates, as is done in the next section.

Also shown in Figure \ref{fig:type} is the fraction of dE candidates that are nucleated. The nucleation fraction appears to vary with host between 0/10 dE candidates around NGC 4631 being nucleated to 7/24 dE candidates around M104. This trend of higher nucleation fraction in larger parent DM halos (we use peak circular speed as a proxy for DM halo mass) in Figure \ref{fig:type} was noted by \citet{trentham2009}. This trend is likely primarily due to the well-established trend of increasing nucleation fraction with the stellar mass of a dwarf galaxy (up to $\log(M_*/M_\odot){\sim}9$) \citep{janssen2019, munoz2015}. The stellar mass of the dwarf satellites in each sample might be different, which is driving the observed trend. Parent halo mass does likely have a smaller, secondary effect on nucleation, but one has to control for the stellar mass of the dwarf satellites, which is out of the scope of the current work. Other than the LG, the nucleation fraction and NSC properties in MW-sized halos is relatively unexplored. This and an analysis of the GCs present in the current dwarf galaxy sample will be in a future paper.

\subsection{Color}
\begin{figure}
\includegraphics[width=0.47\textwidth]{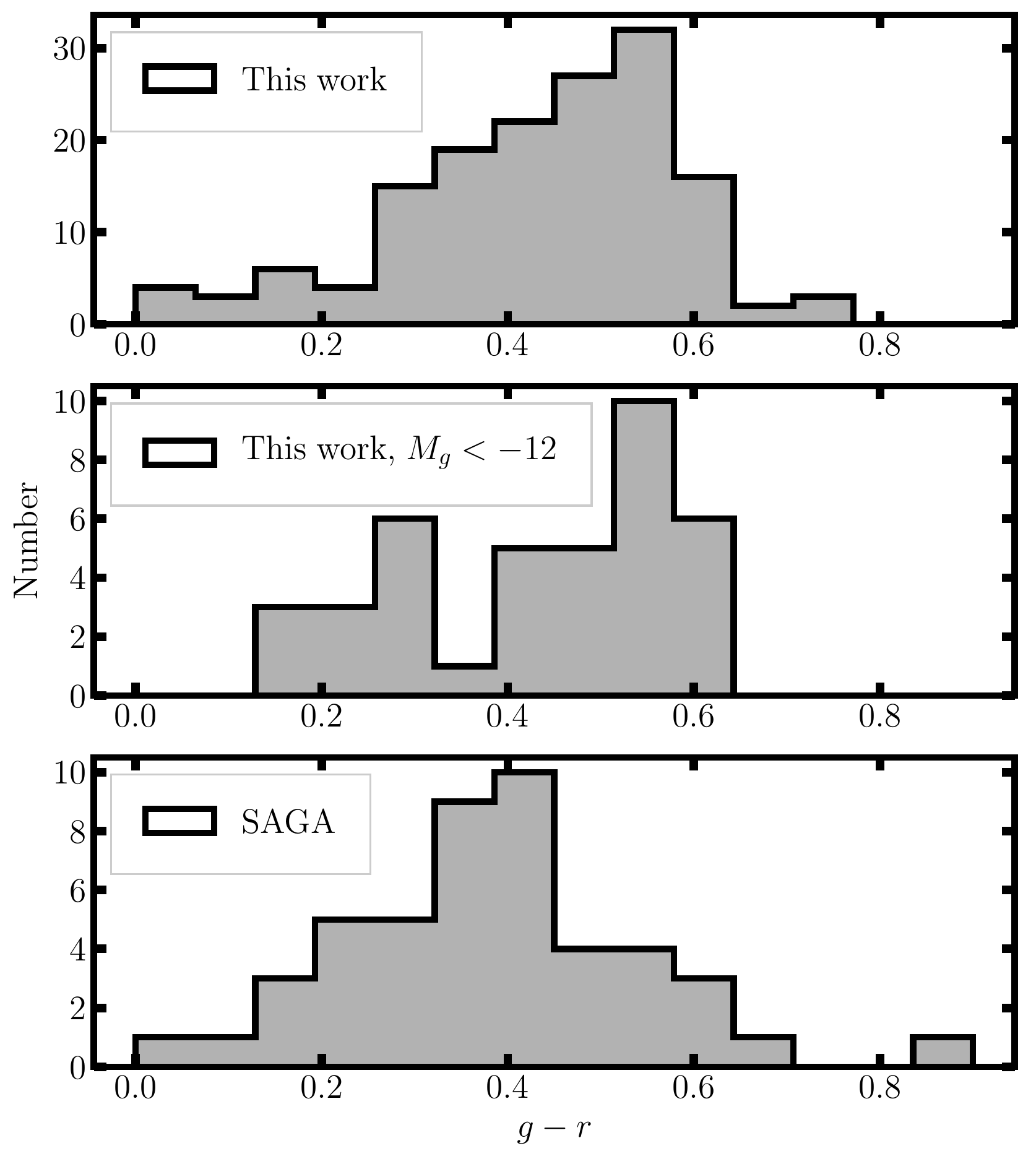}
\caption{The $g-r$ color distribution of our sample of satellite candidates compared with that of the SAGA Survey \citep{geha2017}.}
\label{fig:color}
\end{figure}

As mentioned above, the SAGA Survey \citep{geha2017} cataloged the satellite systems of 8 MW-sized hosts in the distance range $20<D<40$ Mpc down to a completeness magnitude of $M_r<-12.3$ (for a host at 20 Mpc). A weakness of our survey, which only uses photometric data, is that we cannot really quantify the star-forming fraction without follow-up. However, we can compare the color distribution of our satellites with the SAGA satellites. Figure \ref{fig:color} shows this comparison in $g-r$ color distributions. For our satellites around hosts that have $g$ and $i$ band imaging, we convert measured $g-i$ to $g-r$ using a conversion formula derived from MIST isochrones \citep{mist_models}. The CFHT MegaCam $g-r$ color is converted into SDSS to compare with SAGA using:
\beq
(g-r)^{\rm SDSS} = 1.06(g-r)^{\rm CFHT}
\eeq
which comes from the filter transforms on the CFHT website\footnote{\url{http://www.cadc-ccda.hia-iha.nrc-cnrc.gc.ca/en/megapipe/docs/filt.html}} and is an average between the relation expected for the first and second generation MegaCam filters. Our imaging data is roughly split between the two generations of filters. The majority of satellites are fairly red with $g-r{\sim}0.5$ which is consistent with their observed morphology as quenched spheroid systems. Also shown is the distribution of candidate colors restricted to those with $M_g<-12$, to better match our sample with that of the SAGA satellites. Either matched in luminosity or not, our sample appears to have more red $g-r>0.5$ systems than in the SAGA sample\footnote{This trend seems to continue even if we restrict to only the seven hosts in our sample that are most MW-like (NGC 1023, NGC 4258, NGC 4565, NGC 4631, M51, M64) although the statistics get worse without M104.}. A more meaningful comparison will be possible when we use distance information to clean our detected candidates of background contaminants.

\section{Conclusion}
\label{sec:conclusion}
In this paper, we have performed an extensive search for dwarf satellite companions around ten hosts in the Local Volume. Our ten hosts span a range in mass, morphology, and environment. We use archival CFHT/MegaCam imaging to search for dwarf satellites using a semi-automated detection algorithm that is specially designed for LSB dwarf galaxy detection. While many of our hosts have been searched for dwarf companions before, we search all ten using a consistent method which will make comparison of the satellite systems easier. We detect 153 candidates spread across the ten hosts. 93 of these are new detections in the sense that they have not been cataloged as possible satellites before or have archival redshifts. The number of candidates per host ranges from 1 to 33. Some of this is caused by the depth and areal coverage. However, the hosts appear to have large intrinsic scatter. We conduct realistic completeness tests by injecting artificial galaxies before sky subtraction in the raw images. We are generally limited by surface brightness and can detect down to $\mu_{0,r/i}\lesssim26$ mag arcsec$^{-2}$ at $\gtrsim90$\% detection efficiency. Assuming a surface brightness-luminosity relation similar to that of the MW satellites, we can detect dwarfs down to $M_{r/i}{\sim}-8$ to $-9$ in NGC 4258, NGC 4631, M51, M64, and M104. 

We expect that many of these detections will be background contaminants. We plan to conduct an SBF analysis on the detected dwarfs using the same CFHT/MegaCam data used for detection and will present this in a future paper. The ground-based SBF will be able to confirm distances to many of the bright and large satellites. Additionally, the lack of measureable SBF will be able to constrain many candidates to be background. \emph{HST} follow-up will likely be required for some of the faintest and/or smallest candidates that will have ambiguous SBF results.

The survey areas of six of our hosts (M51, M104, NGC 1023, NGC 4631, NGC 4258, and NGC 4565) almost completely cover the inner projected 150 kpc radius volume around the primary. Considering the 3D cone that this encompasses and that satellites are radially concentrated around the host, we survey a significant fraction of the virial volume for these hosts. Prior to this work only six MW analog hosts had been surveyed at this level (MW, M31, M81, CenA, M94, M101). We are therefore roughly doubling the sample of well characterized satellite systems. This much larger sample will improve statistics and allow more detailed tests of structure formation theories.

We survey the two LMC-mass hosts in our sample, NGC 1156 and NGC 5023, out to a projected radius of ${\sim}$50 kpc. This is a significant fraction of the virial radius for these low mass systems ($R_{\rm vir}{\sim}100$ kpc). We find two candidates around NGC 5023 and three around NGC 1156. These candidates are high priority targets for distance follow-up. They are bright $M_r{\sim}-11$ and, if confirmed, would be examples of MW classical-type satellites around isolated LMC-analogs.

Finally, many of our detected dwarfs show interesting properties in their own right. Some systems show a large number of associated point sources, presumably globular clusters. A large fraction of our dwarfs are nucleated. An analysis of the globular cluster systems and nucleation properties will be presented in a future work.

\section*{Acknowledgements}
% -- RLB 
Support for this work was provided by NASA through Hubble Fellowship grant \#51386.01 awarded to R.L.B.by the Space Telescope Science Institute, which is operated by the Association of  Universities for Research in Astronomy, Inc., for NASA, under contract NAS 5-26555. J.P.G. is supported by an NSF Astronomy and Astrophysics Postdoctoral Fellowship under award AST-1801921. J.E.G. is partially supported by the National Science Foundation grant AST-1713828. S.G.C acknowledges support by the National Science Foundation Graduate Research Fellowship Program under Grant No. \#DGE-1656466. Any opinions, findings, and conclusions or recommendations expressed in this material are those of the author(s) and do not necessarily reflect the views of the National Science Foundation.

Based on observations obtained with MegaPrime/MegaCam, a joint project of CFHT and CEA/IRFU, at the Canada-France-Hawaii Telescope (CFHT) which is operated by the National Research Council (NRC) of Canada, the Institut National des Science de l'Univers of the Centre National de la Recherche Scientifique (CNRS) of France, and the University of Hawaii. 

The Pan-STARRS1 Surveys (PS1) and the PS1 public science archive have been made possible through contributions by the Institute for Astronomy, the University of Hawaii, the Pan-STARRS Project Office, the Max-Planck Society and its participating institutes, the Max Planck Institute for Astronomy, Heidelberg and the Max Planck Institute for Extraterrestrial Physics, Garching, The Johns Hopkins University, Durham University, the University of Edinburgh, the Queen's University Belfast, the Harvard-Smithsonian Center for Astrophysics, the Las Cumbres Observatory Global Telescope Network Incorporated, the National Central University of Taiwan, the Space Telescope Science Institute, the National Aeronautics and Space Administration under Grant No. NNX08AR22G issued through the Planetary Science Division of the NASA Science Mission Directorate, the National Science Foundation Grant No. AST-1238877, the University of Maryland, Eotvos Lorand University (ELTE), the Los Alamos National Laboratory, and the Gordon and Betty Moore Foundation.

\software{\texttt{SExtractor} \citep{SExtractor}, \texttt{sep} \citep{sep}, \texttt{Scamp} \citep{scamp}, \texttt{SWarp} \citep{swarp}, \texttt{astropy} \citep{astropy}, \texttt{imfit} \citep{imfit}}

\appendix

\begin{figure*}
\includegraphics[width=\textwidth]{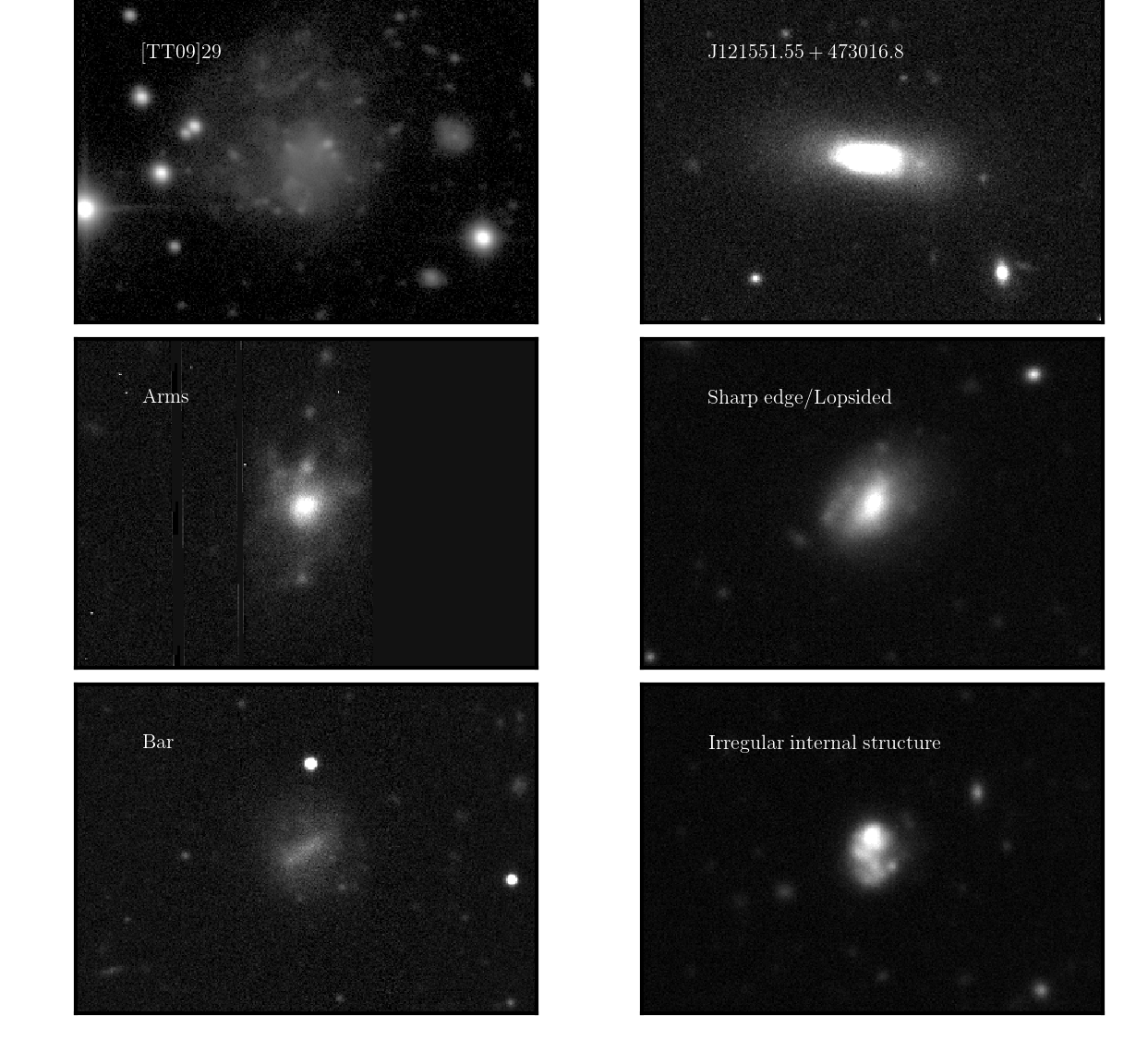}
\caption{Examples of galaxies that were rejected in the visual inspection step. All images are $g$ band cutouts. All images have a linear stretch except for the top left which has an arcsinh stretch. The cutouts are each 1 arcminute wide. The top two are galaxies that were cataloged as possible satellites by previous searches but we reject on the grounds of morphology. Note the sharp ridge and twist in the upper LSB envelope of [TT09]29. J121551.55+473016.8 appears to be a HSB, edge-on system that, given its small size, is likely background. The lower two rows give characteristic examples of galaxies that are rejected along with the reason for rejection.}
\label{fig:rej_ex}
\end{figure*}

\section{Details on Visual Inspection}
\label{app:viz}
Visual inspection plays an important role in the semi-automated dwarf detection algorithm that we use in this paper and we give more details, including examples, on the visual inspection in this section. The goal of the visual inspection is to select out galaxies that could feasibly be at the nearby ($D\lesssim10$ Mpc) distances of our hosts. Intricate morphology, like the presence of spiral arms or a central bar, are only possible in galaxies that are quite large $r_e>20$\arcs~ at the distances of our hosts. Smaller galaxies that exhibit these features are likely background. Small galaxies at the distance of our hosts will be low mass and likely diffuse and fairly regular, like the classical satellites of the MW and M31. 

As mentioned in the main text, there are two examples of candidate satellites that had been cataloged by previous searches and, while recovered by our detection algorithm, were rejected during the visual inspection. Figure \ref{fig:rej_ex} shows these two galaxies along with cutouts of four example rejected galaxies. The rejected galaxies exhibit arms, bars, twists, and/or other irregularities that on a small galaxy indicate the galaxy is fairly distant. As can be seen from the figure, the rejected galaxies are generally small, not especially LSB, and irregular. These galaxies can be compared to the galaxies in Figure \ref{fig:select_dws}, which are more diffuse, regular, and exhibit strong SBF.

\bibliographystyle{aasjournal}
\bibliography{calib}

\end{document}

%% file: ngc1023.tex
\begin{deluxetable*}{ccccccccccccc}
\rotate
\tablecaption{NGC 1023 Dwarf Sample\label{tab:ngc1023}}

\tablewidth{\textwidth}

\tablehead{\colhead{Name} & \colhead{$\alpha$} & \colhead{$\delta$} & \colhead{Type} & \colhead{$m_g$} & \colhead{$m_i$} & \colhead{$g-i$} & \colhead{$r_e$} & \colhead{$\mu_{e,i}$} & \colhead{Star Mask}   & \colhead{D$_{\rm TRGB}$} & \colhead{$cz$} & \colhead{Other Names} \\ 
\colhead{} & \colhead{} & \colhead{} & \colhead{} & \colhead{(mag)} & \colhead{(mag)} & \colhead{} & \colhead{$\prime\prime$} & \colhead{(mag $\prime\prime^{-2}$)} & \colhead{Y/N} & \colhead{(Mpc)}  & \colhead{(km/s)}  & \colhead{} } 

\startdata
dw0233+3852 & 02:33:42.7 & 38:52:20.1 & dE & 18.38$\pm$0.27 & 17.86$\pm$0.24 & 0.52$\pm$0.11 & 14.8$\pm$1.0 & 25.47$\pm$0.17 & Y & & &   \\
dw0234+3800 & 02:34:23.8 & 38:00:32.3 & dI & 19.72$\pm$0.06 & 19.45$\pm$0.06 & 0.27$\pm$0.05 & 5.0$\pm$0.1 & 24.56$\pm$0.07 & N & & &   \\
dw0235+3850 & 02:35:54.2 & 38:50:10.3 & dE & 16.81$\pm$0.15 & 16.21$\pm$0.14 & 0.6$\pm$0.01 & 10.6$\pm$1.7 & 23.7$\pm$0.17 & N & & &   \\
dw0236+3752 & 02:36:12.0 & 37:52:06.2 & dE & 20.05$\pm$0.17 & 19.37$\pm$0.22 & 0.68$\pm$0.05 & 4.0$\pm$0.6 & 24.93$\pm$0.24 & N & & &   \\
IC 239 & 02:36:28.1 & 38:58:08.5 & Scd & 10.7$^{**}$ & 9.9$^{**}$ & 0.76$^{**}$ & 186.0$^{**}$ & 24.1$^{**}$ & N & &903$^a$& [TT09]8$^a$ \\
dw0236+3925 & 02:36:30.8 & 39:25:18.8 & dE & 19.69$\pm$0.22 & 18.93$\pm$0.23 & 0.76$\pm$0.06 & 4.1$\pm$0.5 & 24.65$\pm$0.23 & N & & &   \\
dw0237+3903 & 02:37:09.4 & 39:03:41.0 & dE & 21.15$\pm$0.16 & 20.51$\pm$0.16 & 0.65$\pm$0.07 & 2.3$\pm$0.2 & 24.63$\pm$0.14 & N & & &   \\
dw0237+3855 & 02:37:18.6 & 38:55:59.3 & dE/I,N & 15.18$\pm$0.1 & 14.36$\pm$0.11 & 0.81$\pm$0.02 & 19.2$\pm$1.9 & 23.2$\pm$0.16 & Y & & & [TT09]18$^a$ \\
dw0237+3836 & 02:37:39.4 & 38:36:01.2 & dE & 18.23$\pm$0.17 & 17.58$\pm$0.16 & 0.65$\pm$0.07 & 10.6$\pm$1.3 & 24.66$\pm$0.13 & N & & & [TT09]33$^a$ \\
dw0238+3808 & 02:38:03.0 & 38:08:44.0 & dE & 22.26$\pm$0.23 & 21.9$\pm$0.33 & 0.35$\pm$0.3 & 3.5$\pm$0.5 & 26.6$\pm$0.26 & N & & &   \\
dw0238+3805 & 02:38:41.0 & 38:05:06.5 & dE & 16.7$^{**}$ & 16.2$^{**}$ & 0.57$^{**}$ & 13.0$^{**}$ & 24.2$^{**}$ & Y & & &   \\
dw0239+3926 & 02:39:19.9 & 39:26:02.1 & dE & 17.97$\pm$0.14 & 17.21$\pm$0.14 & 0.75$\pm$0.02 & 25.9$\pm$2.1 & 26.37$\pm$0.16 & N & & & [TT09]38$^a$ \\
dw0239+3910 & 02:39:22.1 & 39:10:22.6 & dE & 22.34$\pm$0.24 & 21.68$\pm$0.16 & 0.66$\pm$0.14 & 4.5$\pm$0.5 & 26.87$\pm$0.18 & N & & &   \\
dw0239+3903 & 02:39:22.5 & 39:03:19.6 & dE,N? & 21.06$\pm$0.49 & 20.36$\pm$0.34 & 0.7$\pm$0.18 & 4.5$\pm$0.9 & 25.74$\pm$0.09 & N & & &   \\
dw0239+3902 & 02:39:47.0 & 39:02:50.4 & dE & 20.62$\pm$0.09 & 19.81$\pm$0.08 & 0.82$\pm$0.02 & 5.3$\pm$0.3 & 25.17$\pm$0.07 & N & & &   \\
dw0239+3824 & 02:39:59.7 & 38:24:04.6 & dE & 21.69$\pm$0.09 & 21.09$\pm$0.04 & 0.6$\pm$0.06 & 3.4$\pm$0.3 & 25.49$\pm$0.19 & N & & & [TT09]61$^a$ \\
dw0240+3844 & 02:40:07.3 & 38:44:56.0 & dI & 20.61$\pm$0.04 & 20.26$\pm$0.06 & 0.34$\pm$0.03 & 3.5$\pm$0.1 & 24.91$\pm$0.05 & N & & &   \\
UGC 2157 & 02:40:25.0 & 38:33:46.9 & Sdm & 13.96$\pm$0.03 & 13.3$\pm$0.03 & 0.66$\pm$0.01 & 39.4$\pm$0.8 & 22.27$\pm$0.02 & N & &488$^a$& [TT09]13$^a$ \\
dw0240+3829 & 02:40:29.5 & 38:29:35.4 & dE & 19.41$\pm$0.08 & 18.97$\pm$0.11 & 0.44$\pm$0.05 & 4.7$\pm$0.3 & 24.51$\pm$0.1 & N & & & [TT09]50$^a$ \\
dw0240+3854 & 02:40:33.0 & 38:54:01.4 & dE & 16.77$\pm$0.03 & 16.34$\pm$0.04 & 0.42$\pm$0.01 & 6.5$\pm$0.1 & 22.12$\pm$0.03 & N & &695$^a$& [TT09]22$^a$ \\
dw0240+3903 & 02:40:37.1 & 39:03:33.6 & dI & 13.0$^{**}$ & 12.2$^{**}$ & 0.76$^{**}$ & 40.6$^{**}$ & 22.5$^{**}$ & N & &743$^a$& [TT09]15$^a$ \\
dw0240+3922 & 02:40:39.6 & 39:22:45.1 & dI & 16.69$\pm$0.07 & 16.43$\pm$0.08 & 0.26$\pm$0.01 & 12.8$\pm$1.1 & 23.7$\pm$0.05 & N & &903$^a$& [TT09]19$^a$ \\
dw0241+3904 & 02:41:00.4 & 39:04:20.6 & dI & 15.92$\pm$0.03 & 15.5$\pm$0.03 & 0.42$\pm$0.01 & 15.5$\pm$0.5 & 23.76$\pm$0.04 & Y & &593$^a$& [TT09]20$^a$ \\
UGC 2165 & 02:41:15.5 & 38:44:38.9 & dE,N & 14.21$\pm$0.04 & 13.32$\pm$0.04 & 0.89$\pm$0.01 & 25.0$\pm$0.8 & 22.34$\pm$0.03 & N & &740$^a$& [TT09]11$^a$ \\
dw0241+3923 & 02:41:16.2 & 39:23:48.4 & dE & 21.62$\pm$0.16 & 20.83$\pm$0.17 & 0.79$\pm$0.09 & 3.6$\pm$0.3 & 25.93$\pm$0.16 & N & & & [TT09]65$^a$ \\
dw0241+3852 & 02:41:20.6 & 38:52:02.2 & dE,N & 21.35$\pm$0.42 & 20.71$\pm$0.36 & 0.64$\pm$0.13 & 6.2$\pm$0.8 & 26.84$\pm$0.22 & N & & &   \\
dw0241+3934 & 02:41:44.4 & 39:34:53.6 & dE & 19.36$\pm$0.08 & 19.2$\pm$0.17 & 0.17$\pm$0.14 & 3.3$\pm$0.3 & 23.39$\pm$0.11 & Y & & &   \\
dw0241+3829 & 02:41:54.2 & 38:29:53.6 & dI & 19.49$\pm$0.13 & 18.85$\pm$0.14 & 0.64$\pm$0.02 & 6.0$\pm$0.6 & 24.67$\pm$0.1 & N & & &   \\
dw0242+3757 & 02:42:22.1 & 37:57:24.5 & dE/I & 22.68$\pm$0.19 & 21.69$\pm$0.1 & 0.99$\pm$0.13 & 1.9$\pm$0.2 & 25.18$\pm$0.14 & N & & &   \\
dw0242+3838 & 02:42:24.6 & 38:38:06.5 & dE & 20.84$\pm$0.11 & 20.4$\pm$0.16 & 0.45$\pm$0.06 & 3.6$\pm$0.2 & 25.32$\pm$0.1 & Y & & &   \\
dw0243+3915 & 02:43:55.0 & 39:15:20.7 & dE & 18.95$\pm$0.13 & 18.21$\pm$0.14 & 0.74$\pm$0.04 & 6.2$\pm$0.9 & 24.76$\pm$0.26 & N & & &   \\
\enddata
\tablecomments{Photometric Properties of the dwarf satellite candidates found in the field around NGC 1023. Galaxies marked with $^{**}$ were either very non-S\'{e}rsic or had some other issue with the fitting and the photometry should be treated with caution. The ``starmask'' column refers to whether the galaxy was found in the visual search of the star mask cutouts, as described in the text.  }
%% General table references marker
\tablerefs{$^{a}$ - \citet{trentham2009}
}
\end{deluxetable*}

%% file: ngc1156.tex
\begin{deluxetable*}{ccccccccccccc}
\rotate
\tablecaption{NGC 1156 Dwarf Sample\label{tab:ngc1156}}

\tablewidth{\textwidth}

\tablehead{\colhead{Name} & \colhead{$\alpha$} & \colhead{$\delta$} & \colhead{Type} & \colhead{$m_g$} & \colhead{$m_i$} & \colhead{$g-i$} & \colhead{$r_e$} & \colhead{$\mu_{e,i}$} & \colhead{Star Mask}   & \colhead{D$_{\rm TRGB}$} & \colhead{$cz$} & \colhead{Other Names} \\ 
\colhead{} & \colhead{} & \colhead{} & \colhead{} & \colhead{(mag)} & \colhead{(mag)} & \colhead{} & \colhead{$\prime\prime$} & \colhead{(mag $\prime\prime^{-2}$)} & \colhead{Y/N} & \colhead{(Mpc)}  & \colhead{(km/s)}  & \colhead{} } 

\startdata
dw0300+2514 & 03:00:17.8 & 25:14:56.0 & dE/I & 19.02$\pm$0.5 & 18.68$\pm$0.41 & 0.33$\pm$0.11 & 5.8$\pm$0.8 & 24.49$\pm$0.21 & Y & & & NGC1156-dw1$^a$ \\
dw0300+2518 & 03:00:27.3 & 25:18:18.3 & dI & 18.82$\pm$0.6 & 18.78$\pm$0.66 & 0.04$\pm$0.15 & 14.0$\pm$1.4 & 26.7$\pm$0.22 & N & & & NGC1156-dw2$^a$ \\
dw0301+2446 & 03:01:32.2 & 24:46:59.4 & dI & 19.04$\pm$0.35 & 18.48$\pm$0.37 & 0.57$\pm$0.21 & 10.5$\pm$1.4 & 25.36$\pm$0.27 & N & & &   \\
\enddata
\tablecomments{Photometric Properties of the dwarf satellite candidates found in the field around NGC 1156. }
%% General table references marker
\tablerefs{$^{a}$ - \citet{karachentsev2015}
}
\end{deluxetable*}

%% file: ngc2903.tex
\begin{deluxetable*}{ccccccccccccc}
\rotate
\tablecaption{NGC 2903 Dwarf Sample\label{tab:ngc2903}}

\tablewidth{\textwidth}

\tablehead{\colhead{Name} & \colhead{$\alpha$} & \colhead{$\delta$} & \colhead{Type} & \colhead{$m_g$} & \colhead{$m_i$} & \colhead{$g-i$} & \colhead{$r_e$} & \colhead{$\mu_{e,i}$} & \colhead{Star Mask}   & \colhead{D$_{\rm TRGB}$} & \colhead{$cz$} & \colhead{Other Names} \\ 
\colhead{} & \colhead{} & \colhead{} & \colhead{} & \colhead{(mag)} & \colhead{(mag)} & \colhead{} & \colhead{$\prime\prime$} & \colhead{(mag $\prime\prime^{-2}$)} & \colhead{Y/N} & \colhead{(Mpc)}  & \colhead{(km/s)}  & \colhead{} } 

\startdata
dw0930+2143 & 09:30:40.0 & 21:43:27.1 & dI & 18.66$\pm$0.08 & 18.42$\pm$0.07 & 0.24$\pm$0.03 & 7.7$\pm$0.6 & 24.49$\pm$0.11 & N & &582$^a$& N2903-HI-1$^a$ \\
UGC 5086 & 09:32:48.8 & 21:27:56.2 & dE,N & 15.73$\pm$0.08 & 15.17$\pm$0.06 & 0.56$\pm$0.03 & 16.9$\pm$0.8 & 23.78$\pm$0.06 & N &7.4$^{\dagger}$$^b$&491$^c$&   \\
dw0933+2114 & 09:33:28.5 & 21:14:00.1 & dE & 21.41$\pm$0.49 & 20.87$\pm$0.39 & 0.54$\pm$0.1 & 4.8$\pm$0.6 & 26.78$\pm$0.1 & N & & &   \\
dw0934+2204 & 09:34:22.0 & 22:04:53.9 & dE & 19.37$\pm$0.13 & 19.08$\pm$0.13 & 0.29$\pm$0.02 & 3.8$\pm$0.6 & 24.48$\pm$0.22 & N & & &   \\
\enddata
\tablecomments{Photometric Properties of the dwarf satellite candidates found in the field around NGC 2903. \\
$^{\dagger}$ - Not a TRGB distance but a brightest stars distance.}
%% General table references marker
\tablerefs{$^{a}$ - \citet{irwin2009}
$^{b}$ - \citet{makarova1998}
$^{c}$ - \citet{springob2005}
}
\end{deluxetable*}

%% file: ngc4258.tex
\begin{deluxetable*}{ccccccccccccc}
\tablecaption{NGC 4258 Dwarf Sample\label{tab:ngc4258}}

\tablewidth{\textwidth}
\rotate

\tablehead{\colhead{Name} & \colhead{$\alpha$} & \colhead{$\delta$} & \colhead{Type} & \colhead{$m_g$} & \colhead{$m_i$} & \colhead{$g-i$} & \colhead{$r_e$} & \colhead{$\mu_{e,i}$} & \colhead{Star Mask}   & \colhead{D$_{\rm TRGB}$} & \colhead{$cz$} & \colhead{Other Names} \\ 
\colhead{} & \colhead{} & \colhead{} & \colhead{} & \colhead{(mag)} & \colhead{(mag)} & \colhead{} & \colhead{$\prime\prime$} & \colhead{(mag $\prime\prime^{-2}$)} & \colhead{Y/N} & \colhead{(Mpc)}  & \colhead{(km/s)}  & \colhead{} }

\startdata
dw1214+4726 & 12:14:05.0 & 47:26:08.2 & dE & 18.96$\pm$0.19 & 18.55$\pm$0.14 & 0.42$\pm$0.06 & 6.1$\pm$0.7 & 24.63$\pm$0.17 & N & & & S1$^a$ \\
dw1214+4621 & 12:14:40.0 & 46:21:12.6 & dI & 18.27$\pm$0.16 & 17.95$\pm$0.15 & 0.32$\pm$0.05 & 5.4$\pm$0.4 & 23.48$\pm$0.18 & N & &583$^c$& 480$^c$ \\
dw1214+4743 & 12:14:52.8 & 47:43:17.6 & dE & 21.06$\pm$0.25 & 20.64$\pm$0.24 & 0.42$\pm$0.13 & 5.7$\pm$0.8 & 26.57$\pm$0.24 & N & & &   \\
dw1216+4709 & 12:16:37.4 & 47:09:11.9 & dE & 22.27$\pm$0.2 & 21.8$\pm$0.22 & 0.47$\pm$0.11 & 3.2$\pm$0.3 & 26.23$\pm$0.26 & N & & & P5$^a$ \\
dw1217+4639 & 12:17:00.8 & 46:39:10.1 & dE & 19.45$\pm$0.04 & 19.03$\pm$0.04 & 0.42$\pm$0.02 & 3.9$\pm$0.1 & 24.3$\pm$0.07 & N & & &   \\
dw1217+4703 & 12:17:09.4 & 47:03:52.2 & dE,N? & 19.24$\pm$0.22 & 18.69$\pm$0.2 & 0.55$\pm$0.04 & 6.8$\pm$0.7 & 24.7$\pm$0.06 & N &$>$9.7$^b$& & BTS 109$^g$, DF3$^b$, S6$^a$ \\
dw1217+4759 & 12:17:32.0 & 47:59:42.8 & dI & 15.48$\pm$0.04 & 15.23$\pm$0.03 & 0.25$\pm$0.01 & 7.7$\pm$0.2 & 21.78$\pm$0.05 & N & &704$^f$& 080$^c$ \\
dw1217+4747 & 12:17:35.9 & 47:47:47.5 & dI & 21.59$\pm$0.17 & 21.43$\pm$0.22 & 0.16$\pm$0.08 & 3.9$\pm$0.6 & 26.6$\pm$0.18 & N & & &   \\
NGC 4248 & 12:17:50.2 & 47:24:33.4 & Im & 12.72$\pm$0.02 & 12.22$\pm$0.02 & 0.5$\pm$0.01 & 52.3$\pm$0.5 & 22.28$\pm$0.02 & N & &527$^f$& S2$^a$ \\
dw1217+4656 & 12:17:59.4 & 46:56:34.2 & dI & 18.05$\pm$0.02 & 17.96$\pm$0.02 & 0.09$\pm$0.01 & 3.9$\pm$0.04 & 23.06$\pm$0.03 & N & & &   \\
dw1218+4623 & 12:18:02.6 & 46:23:05.4 & dE & 19.99$\pm$0.44 & 19.56$\pm$0.44 & 0.43$\pm$0.09 & 18.6$\pm$3.2 & 27.95$\pm$0.15 & N & & &   \\
LVJ1218+4655 & 12:18:11.2 & 46:55:02.0 & dI & 16.49$\pm$0.02 & 16.27$\pm$0.03 & 0.22$\pm$0.01 & 16.2$\pm$0.6 & 23.44$\pm$0.07 & N &8.28&387$^f$& S5$^a$ \\
dw1218+4748 & 12:18:18.8 & 47:48:17.5 & dE & 21.33$\pm$0.18 & 20.77$\pm$0.21 & 0.56$\pm$0.08 & 4.2$\pm$0.5 & 26.23$\pm$0.14 & N & & & P1$^a$ \\
dw1218+4801 & 12:18:51.2 & 48:01:28.4 & dE & 23.35$\pm$0.18 & 22.94$\pm$0.17 & 0.4$\pm$0.14 & 2.2$\pm$0.3 & 26.74$\pm$0.27 & N & & &   \\
dw1219+4743 & 12:19:06.2 & 47:43:49.3 & dE & 18.53$\pm$0.16 & 18.13$\pm$0.17 & 0.4$\pm$0.04 & 10.4$\pm$1.3 & 25.58$\pm$0.1 & N &7.3$^b$& & S6$^a$, DF6$^b$, KK132$^d$ \\
UGC 7356 & 12:19:09.0 & 47:05:23.9 & dE,N & 15.28$\pm$0.09 & 14.76$\pm$0.1 & 0.52$\pm$0.05 & 25.7$\pm$1.7 & 23.84$\pm$0.07 & N &7.28&135$^f$& KDG 101, S7$^a$ \\
dw1219+4921 & 12:19:17.8 & 49:21:20.9 & dE & 19.93$\pm$0.1 & 19.46$\pm$0.12 & 0.47$\pm$0.04 & 4.3$\pm$0.9 & 24.75$\pm$0.23 & N & & &   \\
dw1219+4718 & 12:19:27.3 & 47:18:44.5 & dI & 18.47$\pm$0.08 & 18.14$\pm$0.07 & 0.33$\pm$0.02 & 4.5$\pm$0.2 & 23.65$\pm$0.09 & N & & & 358$^c$ \\
dw1219+4727 & 12:19:33.2 & 47:27:05.4 & dE,N? & 16.87$\pm$0.1 & 16.4$\pm$0.1 & 0.47$\pm$0.02 & 14.4$\pm$1.5 & 24.85$\pm$0.13 & N & &785$^f$& BTS 118$^g$, KK 134$^d$, S8$^a$ \\
dw1219+4705 & 12:19:36.0 & 47:05:35.8 & dE,N & 18.35$\pm$0.12 & 17.84$\pm$0.12 & 0.51$\pm$0.03 & 9.4$\pm$0.9 & 25.27$\pm$0.09 & N &$>$10.1$^b$& & d1219+4705$^e$, S9$^a$, DF2$^b$ \\
dw1219+4939 & 12:19:50.1 & 49:39:26.7 & dE & 20.05$\pm$0.11 & 19.6$\pm$0.13 & 0.45$\pm$0.03 & 3.3$\pm$0.3 & 24.68$\pm$0.14 & N & & &   \\
dw1220+4919 & 12:20:05.0 & 49:19:18.0 & dE & 19.78$\pm$0.04 & 19.44$\pm$0.06 & 0.35$\pm$0.05 & 2.8$\pm$0.1 & 24.05$\pm$0.12 & N & & &   \\
dw1220+4922 & 12:20:14.4 & 49:22:51.4 & dE & 19.99$\pm$0.04 & 19.7$\pm$0.08 & 0.29$\pm$0.06 & 4.8$\pm$0.2 & 25.34$\pm$0.14 & N & & &   \\
UGC 7392 & 12:20:17.5 & 48:08:11.9 & dI & 15.86$\pm$0.02 & 15.6$\pm$0.03 & 0.26$\pm$0.01 & 16.0$\pm$0.3 & 22.54$\pm$0.02 & N & &806$^f$& S10$^a$ \\
dw1220+4729 & 12:20:30.2 & 47:29:26.7 & dE & 20.12$\pm$0.35 & 19.85$\pm$0.27 & 0.27$\pm$0.12 & 13.7$\pm$2.0 & 27.6$\pm$0.2 & N & & & S11$^a$ \\
dw1220+4700 & 12:20:40.2 & 47:00:03.1 & dE & 17.07$\pm$0.06 & 16.54$\pm$0.06 & 0.54$\pm$0.01 & 14.8$\pm$0.8 & 24.93$\pm$0.06 & N &$>$12.5$^b$& & KK 136$^d$, S12$^a$, DF1$^b$ \\
UGC7401 & 12:20:49.1 & 47:49:44.2 & Im & 15.19$\pm$0.03 & 14.89$\pm$0.04 & 0.3$\pm$0.01 & 29.8$\pm$0.6 & 24.31$\pm$0.02 & N & &759$^f$& S13$^a$ \\
dw1220+4649 & 12:20:54.9 & 46:49:48.4 & dE & 18.82$\pm$0.13 & 18.33$\pm$0.1 & 0.49$\pm$0.05 & 11.5$\pm$0.7 & 25.97$\pm$0.05 & N & & & d1220+4649$^e$, S14$^a$ \\
dw1220+4748 & 12:20:55.8 & 47:48:59.4 & dE & 21.98$\pm$0.38 & 21.62$\pm$0.31 & 0.36$\pm$0.13 & 4.6$\pm$0.9 & 26.64$\pm$0.25 & N & & &   \\
dw1222+4755 & 12:22:54.9 & 47:55:42.6 & dE & 18.06$\pm$0.02 & 17.63$\pm$0.03 & 0.43$\pm$0.02 & 4.9$\pm$0.1 & 23.54$\pm$0.05 & N & & &   \\
dw1223+4848 & 12:23:12.8 & 48:48:56.4 & dE & 20.9$\pm$0.22 & 20.36$\pm$0.16 & 0.54$\pm$0.07 & 4.7$\pm$0.6 & 25.99$\pm$0.18 & Y & & &   \\
dw1223+4739 & 12:23:46.2 & 47:39:32.7 & dE,N & 18.02$\pm$0.09 & 17.57$\pm$0.08 & 0.45$\pm$0.04 & 17.2$\pm$2.3 & 25.7$\pm$0.08 & N & & & S16$^a$ \\
dw1223+4920 & 12:23:55.8 & 49:20:15.2 & dE & 18.37$\pm$0.07 & 17.99$\pm$0.08 & 0.39$\pm$0.02 & 6.7$\pm$0.4 & 24.6$\pm$0.08 & N & & &   \\
\enddata
\tablecomments{Photometric Properties of the dwarf satellite candidates found in the field around NGC 4258. }
%% General table references marker
\tablerefs{$^{a}$ - \citet{kim2011}
$^{b}$ - \citet{cohen2018}
$^{c}$ - \citet{spencer2014}
$^{d}$ - \citet{kk98}
$^{e}$ - \citet{kkh07}
$^{f}$ - SIMBAD
$^{g}$ - \citet{binggeli1990}
}
\end{deluxetable*}

%% file: ngc4565.tex
\begin{deluxetable*}{ccccccccccccc}
\rotate
\tablecaption{NGC 4565 Dwarf Sample\label{tab:ngc4565}}

\tablewidth{\textwidth}

\tablehead{\colhead{Name} & \colhead{$\alpha$} & \colhead{$\delta$} & \colhead{Type} & \colhead{$m_g$} & \colhead{$m_i$} & \colhead{$g-i$} & \colhead{$r_e$} & \colhead{$\mu_{e,i}$} & \colhead{Star Mask}   & \colhead{D$_{\rm TRGB}$} & \colhead{$cz$} & \colhead{Other Names} \\ 
\colhead{} & \colhead{} & \colhead{} & \colhead{} & \colhead{(mag)} & \colhead{(mag)} & \colhead{} & \colhead{$\prime\prime$} & \colhead{(mag $\prime\prime^{-2}$)} & \colhead{Y/N} & \colhead{(Mpc)}  & \colhead{(km/s)}  & \colhead{} }

\startdata
dw1233+2535 & 12:33:11.0 & 25:35:55.2 & dE & 18.65$\pm$0.07 & 18.25$\pm$0.08 & 0.4$\pm$0.01 & 4.7$\pm$0.2 & 23.63$\pm$0.07 & N & & &   \\
dw1233+2543 & 12:33:18.4 & 25:43:35.1 & dE & 20.57$\pm$0.1 & 20.23$\pm$0.11 & 0.34$\pm$0.03 & 4.1$\pm$0.2 & 25.49$\pm$0.1 & N & & &   \\
dw1234+2531 & 12:34:24.2 & 25:31:20.2 & dE,N & 16.65$\pm$0.03 & 16.15$\pm$0.02 & 0.5$\pm$0.01 & 19.9$\pm$0.5 & 24.31$\pm$0.04 & N & &501$^a$&   \\
dw1234+2627 & 12:34:25.0 & 26:27:16.4 & dE & 21.85$\pm$0.26 & 21.39$\pm$0.26 & 0.46$\pm$0.09 & 3.7$\pm$0.6 & 26.76$\pm$0.26 & N & & &   \\
dw1234+2618 & 12:34:57.6 & 26:18:50.8 & dI & 20.25$\pm$0.06 & 19.93$\pm$0.05 & 0.32$\pm$0.03 & 4.6$\pm$0.2 & 25.6$\pm$0.11 & N & & &   \\
dw1235+2616 & 12:35:22.3 & 26:16:14.2 & dE & 20.54$\pm$0.13 & 20.02$\pm$0.15 & 0.51$\pm$0.03 & 4.5$\pm$0.9 & 25.6$\pm$0.23 & N & & &   \\
NGC 4562 & 12:35:34.7 & 25:51:01.3 & Im & 13.5$\pm$0.01 & 13.05$\pm$0.01 & 0.45$\pm$0.01 & 35.4$\pm$0.1 & 22.19$\pm$0.01 & N & &1353$^a$&   \\
dw1235+2534 & 12:35:37.5 & 25:34:12.2 & dE & 21.93$\pm$0.21 & 21.58$\pm$0.22 & 0.36$\pm$0.09 & 4.3$\pm$0.8 & 26.65$\pm$0.28 & N & & &   \\
dw1235+2637 & 12:35:42.2 & 26:37:14.7 & dE & 21.68$\pm$0.31 & 21.62$\pm$0.43 & 0.06$\pm$0.25 & 6.7$\pm$1.7 & 27.5$\pm$0.26 & N & & &   \\
dw1235+2609 & 12:35:55.2 & 26:09:55.4 & dE & 22.74$\pm$0.21 & 22.35$\pm$0.25 & 0.39$\pm$0.08 & 3.0$\pm$0.5 & 26.97$\pm$0.26 & N & & &   \\
dw1235+2606 & 12:35:56.4 & 26:06:52.3 & dI & 19.06$\pm$0.3 & 18.97$\pm$0.33 & 0.09$\pm$0.16 & 27.5$\pm$4.1 & 27.99$\pm$0.24 & N & & &   \\
dw1236+2616 & 12:36:05.9 & 26:16:25.7 & dE & 22.88$\pm$0.15 & 22.4$\pm$0.12 & 0.48$\pm$0.06 & 3.0$\pm$0.2 & 26.35$\pm$0.18 & N & & &   \\
dw1236+2605 & 12:36:20.0 & 26:05:03.5 & dI & 16.59$\pm$0.01 & 16.43$\pm$0.01 & 0.15$\pm$0.01 & 8.5$\pm$0.1 & 23.07$\pm$0.02 & N & &1202$^a$&   \\
dw1236+2603 & 12:36:25.2 & 26:03:18.7 & dE & 21.45$\pm$0.19 & 21.17$\pm$0.2 & 0.28$\pm$0.07 & 4.5$\pm$0.7 & 26.64$\pm$0.19 & N & & &   \\
dw1236+2634 & 12:36:58.6 & 26:34:42.8 & dE & 21.17$\pm$0.18 & 20.68$\pm$0.19 & 0.5$\pm$0.04 & 4.7$\pm$0.5 & 26.21$\pm$0.15 & N & & &   \\
dw1237+2602 & 12:37:01.2 & 26:02:09.6 & dE,N & 17.96$\pm$0.06 & 17.58$\pm$0.07 & 0.38$\pm$0.01 & 8.4$\pm$0.6 & 24.31$\pm$0.13 & N & & &   \\
dw1237+2605 & 12:37:26.8 & 26:05:08.7 & dE,N & 19.77$\pm$0.32 & 19.36$\pm$0.35 & 0.41$\pm$0.07 & 13.2$\pm$3.4 & 26.99$\pm$0.26 & N & & &   \\
dw1237+2637 & 12:37:42.8 & 26:37:27.6 & dE & 20.23$\pm$0.08 & 19.71$\pm$0.1 & 0.52$\pm$0.06 & 4.9$\pm$0.6 & 25.13$\pm$0.1 & N & & &   \\
dw1237+2631 & 12:37:54.6 & 26:31:08.0 & dE & 22.51$\pm$0.36 & 22.08$\pm$0.32 & 0.43$\pm$0.09 & 2.7$\pm$0.3 & 25.88$\pm$0.11 & N & & &   \\
dw1238+2610 & 12:38:39.6 & 26:10:01.0 & dE & 21.98$\pm$0.29 & 21.55$\pm$0.3 & 0.44$\pm$0.09 & 4.7$\pm$0.5 & 26.9$\pm$0.23 & N & & &   \\
dw1238+2536 & 12:38:54.6 & 25:36:56.2 & dE & 20.92$\pm$0.07 & 20.48$\pm$0.09 & 0.44$\pm$0.08 & 3.5$\pm$0.2 & 25.14$\pm$0.11 & N & & &   \\
\enddata
\tablecomments{Photometric Properties of the dwarf satellite candidates found in the field around NGC 4565. }
%% General table references marker
\tablerefs{$^{a}$ - SIMBAD
}
\end{deluxetable*}

%% file: ngc4631.tex
\begin{deluxetable*}{ccccccccccccc}
\rotate
\tablecaption{NGC 4631 Dwarf Sample\label{tab:ngc4631}}

\tablewidth{\textwidth}

\tablehead{\colhead{Name} & \colhead{$\alpha$} & \colhead{$\delta$} & \colhead{Type} & \colhead{$m_g$} & \colhead{$m_i$} & \colhead{$g-i$} & \colhead{$r_e$} & \colhead{$\mu_{e,i}$} & \colhead{Star Mask}   & \colhead{D$_{\rm TRGB}$} & \colhead{$cz$} & \colhead{Other Names} \\ 
\colhead{} & \colhead{} & \colhead{} & \colhead{} & \colhead{(mag)} & \colhead{(mag)} & \colhead{} & \colhead{$\prime\prime$} & \colhead{(mag $\prime\prime^{-2}$)} & \colhead{Y/N} & \colhead{(Mpc)}  & \colhead{(km/s)}  & \colhead{} } 

\startdata
UGCA 292 & 12:38:40.4 & 32:45:52.7 & dI & 15.83$\pm$0.12 & 15.68$\pm$0.12 & 0.16$\pm$0.05 & 34.2$\pm$4.7 & 25.4$\pm$0.12 & N &3.6$^d$&307$^a$& CVn I dwA \\
dw1239+3230 & 12:39:05.0 & 32:30:16.5 & dE/I & 19.09$\pm$0.09 & 18.8$\pm$0.08 & 0.29$\pm$0.03 & 8.0$\pm$0.5 & 25.78$\pm$0.06 & N & & &   \\
dw1239+3251 & 12:39:19.6 & 32:51:39.3 & dE & 20.03$\pm$0.31 & 19.46$\pm$0.27 & 0.58$\pm$0.11 & 13.7$\pm$2.8 & 27.25$\pm$0.15 & N & & &   \\
dw1240+3239 & 12:40:09.9 & 32:39:31.6 & dI & 16.0$\pm$0.03 & 15.77$\pm$0.03 & 0.22$\pm$0.02 & 18.3$\pm$0.4 & 23.91$\pm$0.04 & N & &776$^a$& HSC-7$^c$ \\
dw1240+3216 & 12:40:53.0 & 32:16:55.9 & dE & 19.0$\pm$0.1 & 18.51$\pm$0.09 & 0.49$\pm$0.05 & 8.7$\pm$0.6 & 25.14$\pm$0.08 & N & & & HSC-9$^c$ \\
dw1240+3247 & 12:40:58.5 & 32:47:25.0 & dE & 16.07$\pm$0.64 & 15.5$\pm$0.42 & 0.57$\pm$0.33 & 71.1$\pm$19.1 & 26.66$\pm$0.26 & N & & & dw1$^f$ \\
dw1241+3251 & 12:41:47.1 & 32:51:27.3 & dI & 15.7$\pm$0.05 & 15.56$\pm$0.07 & 0.13$\pm$0.02 & 18.0$\pm$0.9 & 23.77$\pm$0.04 & N & &696$^a$& HSC-8$^c$ \\
NGC 4627 & 12:41:59.7 & 32:34:26.2 & dE? & 12.8$^{**}$ & 12.5$^{**}$ & 0.36$^{**}$ & 27.1$^{**}$ & 22.0$^{**}$ & N & &813$^b$&   \\
dw1242+3224 & 12:42:01.6 & 32:24:06.4 & dI & 20.1$\pm$0.07 & 20.0$\pm$0.15 & 0.1$\pm$0.11 & 3.7$\pm$0.1 & 24.54$\pm$0.14 & N & & & HSC-11$^c$ \\
dw1242+3237 & 12:42:06.2 & 32:37:18.7 & dE & 18.88$\pm$0.43 & 18.48$\pm$0.45 & 0.4$\pm$0.17 & 18.4$\pm$2.8 & 27.06$\pm$0.19 & N & & & HSC-2$^c$, DGSAT-2$^e$, dw2$^f$ \\
dw1242+3231 & 12:42:28.4 & 32:31:51.1 & dI? & 17.07$\pm$0.18 & 16.65$\pm$0.18 & 0.42$\pm$0.01 & 11.2$\pm$1.8 & 24.13$\pm$0.18 & N & & &   \\
dw1242+3158 & 12:42:31.4 & 31:58:09.2 & dE & 19.12$\pm$0.1 & 18.64$\pm$0.08 & 0.49$\pm$0.05 & 8.2$\pm$0.6 & 25.35$\pm$0.13 & N & & & HSC-10$^c$ \\
dw1242+3227 & 12:42:53.2 & 32:27:19.4 & dE & 19.38$\pm$0.21 & 19.34$\pm$0.38 & 0.04$\pm$0.19 & 12.3$\pm$1.9 & 27.08$\pm$0.13 & N & & & HSC-1$^c$, DGSAT-1$^e$, dw3$^f$ \\
dw1243+3229 & 12:43:07.0 & 32:29:27.3 & dI & 14.97$\pm$0.04 & 14.71$\pm$0.04 & 0.25$\pm$0.01 & 9.2$\pm$0.2 & 21.89$\pm$0.06 & N & &892$^a$& HSC-12$^c$ \\
dw1243+3228 & 12:43:24.8 & 32:28:55.3 & dE & 16.73$\pm$0.03 & 16.29$\pm$0.03 & 0.44$\pm$0.01 & 16.6$\pm$0.3 & 24.3$\pm$0.03 & N & &665$^b$& HSC-6$^c$ \\
dw1243+3232 & 12:43:44.8 & 32:32:03.6 & dE & 18.64$\pm$0.05 & 18.2$\pm$0.04 & 0.44$\pm$0.01 & 10.3$\pm$0.3 & 25.4$\pm$0.05 & N & & & HSC-5$^c$ \\
\enddata
\tablecomments{Photometric Properties of the dwarf satellite candidates found in the field around NGC 4631. Galaxies marked with $^{**}$ were either very non-S\'{e}rsic or had some other issue with the fitting and the photometry should be treated with caution.}
%% General table references marker
\tablerefs{$^{a}$ - \citet{cairns}
$^{b}$ - SIMBAD
$^{c}$ - \citet{tanaka2017}
$^{d}$ - \citet{dalcanton2009}
$^{e}$ - \citet{javanmardi_m101}
$^{f}$ - \citet{karachentsev2015}
}
\end{deluxetable*}

%% file: ngc5023.tex
\begin{deluxetable*}{ccccccccccccc}
\rotate
\tablecaption{NGC 5023 Dwarf Sample\label{tab:ngc5023}}

\tablewidth{\textwidth}

\tablehead{\colhead{Name} & \colhead{$\alpha$} & \colhead{$\delta$} & \colhead{Type} & \colhead{$m_g$} & \colhead{$m_i$} & \colhead{$g-i$} & \colhead{$r_e$} & \colhead{$\mu_{e,i}$} & \colhead{Star Mask}   & \colhead{D$_{\rm TRGB}$} & \colhead{$cz$} & \colhead{Other Names} \\ 
\colhead{} & \colhead{} & \colhead{} & \colhead{} & \colhead{(mag)} & \colhead{(mag)} & \colhead{} & \colhead{$\prime\prime$} & \colhead{(mag $\prime\prime^{-2}$)} & \colhead{Y/N} & \colhead{(Mpc)}  & \colhead{(km/s)}  & \colhead{} } 

\startdata
dw1310+4358 & 13:10:59.97 & 43:58:48.6 & dE & 21.68$\pm$0.15 & 20.8$\pm$0.17 & 0.88$\pm$0.09 & 4.0$\pm$0.5 & 26.04$\pm$0.21 & N & & &   \\
dw1314+4420 & 13:14:34.4 & 44:20:03.1 & dE & 22.39$\pm$0.16 & 21.81$\pm$0.27 & 0.58$\pm$0.18 & 3.1$\pm$0.3 & 26.08$\pm$0.23 & N & & &   \\
\enddata
\tablecomments{Photometric Properties of the dwarf satellite candidates found in the field around NGC 5023. }
%% General table references marker
\tablerefs{$^{a}$ - \citet{trentham2009}
}
\end{deluxetable*}

%% file: m51.tex
\begin{deluxetable*}{ccccccccccccc}
\rotate
\tablecaption{M51 Dwarf Sample\label{tab:m51}}

\tablewidth{\textwidth}

\tablehead{\colhead{Name} & \colhead{$\alpha$} & \colhead{$\delta$} & \colhead{Type} & \colhead{$m_g$} & \colhead{$m_i$} & \colhead{$g-i$} & \colhead{$r_e$} & \colhead{$\mu_{e,i}$} & \colhead{Star Mask}   & \colhead{D$_{\rm TRGB}$} & \colhead{$cz$} & \colhead{Other Names} \\ 
\colhead{} & \colhead{} & \colhead{} & \colhead{} & \colhead{(mag)} & \colhead{(mag)} & \colhead{} & \colhead{$\prime\prime$} & \colhead{(mag $\prime\prime^{-2}$)} & \colhead{Y/N} & \colhead{(Mpc)}  & \colhead{(km/s)}  & \colhead{} } 

\startdata
dw1327+4637 & 13:27:10.6 & 46:37:56.3 & dE/I & 21.37$\pm$0.24 & 20.82$\pm$0.21 & 0.55$\pm$0.11 & 4.7$\pm$1.0 & 26.64$\pm$0.17 & N & & &   \\
dw1327+4654 & 13:27:37.5 & 46:54:54.2 & dE & 19.01$\pm$0.05 & 18.67$\pm$0.03 & 0.34$\pm$0.02 & 3.5$\pm$0.1 & 24.02$\pm$0.06 & N & & &   \\
dw1327+4626 & 13:27:53.1 & 46:26:28.6 & dE/I & 20.75$\pm$0.2 & 20.39$\pm$0.22 & 0.36$\pm$0.03 & 3.9$\pm$0.8 & 25.51$\pm$0.14 & N & & &   \\
dw1328+4718 & 13:28:21.9 & 47:18:14.6 & dI & 19.01$\pm$0.04 & 18.7$\pm$0.05 & 0.31$\pm$0.02 & 6.3$\pm$0.2 & 24.15$\pm$0.07 & N & & &   \\
dw1328+4703 & 13:28:24.7 & 47:03:54.8 & dE & 20.4$\pm$0.11 & 19.82$\pm$0.11 & 0.58$\pm$0.04 & 6.7$\pm$0.4 & 25.98$\pm$0.12 & N & & &   \\
dw1329+4634 & 13:29:51.2 & 46:34:57.1 & dE,N & 17.33$\pm$0.04 & 16.82$\pm$0.04 & 0.51$\pm$0.01 & 8.1$\pm$0.3 & 24.06$\pm$0.05 & N & &2676$^a$&   \\
dw1329+4622 & 13:29:53.7 & 46:22:19.2 & dE & 19.52$\pm$0.15 & 19.01$\pm$0.16 & 0.51$\pm$0.03 & 7.1$\pm$0.8 & 25.83$\pm$0.16 & N & &5600$^c$& PG 1327+4637$^c$ \\
dw1330+4708 & 13:30:32.2 & 47:08:15.6 & dE & 19.68$\pm$0.06 & 19.23$\pm$0.07 & 0.45$\pm$0.02 & 6.8$\pm$0.8 & 25.85$\pm$0.24 & N & &109208$^a$&   \\
dw1330+4731 & 13:30:33.9 & 47:31:33.1 & dE & 20.0$\pm$0.15 & 19.65$\pm$0.17 & 0.36$\pm$0.11 & 12.4$\pm$1.2 & 27.14$\pm$0.18 & N & & &   \\
dw1330+4720 & 13:30:44.4 & 47:20:43.5 & dE & 20.03$\pm$0.07 & 19.53$\pm$0.09 & 0.5$\pm$0.04 & 4.7$\pm$0.4 & 25.36$\pm$0.18 & N & & &   \\
dw1331+4654 & 13:31:08.2 & 46:54:27.8 & dE & 22.22$\pm$0.08 & 21.68$\pm$0.07 & 0.54$\pm$0.08 & 3.1$\pm$0.1 & 26.52$\pm$0.19 & N & & &   \\
dw1331+4648 & 13:31:11.6 & 46:48:57.4 & dE & 20.65$\pm$0.17 & 20.22$\pm$0.17 & 0.43$\pm$0.03 & 6.9$\pm$1.1 & 27.02$\pm$0.29 & N & & &   \\
dw1332+4703 & 13:32:45.9 & 47:03:02.6 & dE & 19.6$\pm$0.06 & 19.06$\pm$0.07 & 0.54$\pm$0.01 & 3.6$\pm$0.1 & 24.08$\pm$0.09 & N & & &   \\
dw1333+4725 & 13:33:45.2 & 47:25:35.8 & dI & 17.76$\pm$0.11 & 17.25$\pm$0.11 & 0.51$\pm$0.02 & 4.7$\pm$0.5 & 22.87$\pm$0.13 & N & & &   \\
NGC 5229 & 13:34:03.0 & 47:54:49.8 & dI & 13.73$\pm$0.01 & 13.33$\pm$0.01 & 0.39$\pm$0.01 & 37.0$\pm$0.1 & 21.94$\pm$0.01 & N &5.13$^{\dagger}$$^b$&364$^a$&   \\
\enddata
\tablecomments{Photometric Properties of the dwarf satellite candidates found in the field around M51. 
$^\dagger$ - Not actually a TRGB distance but a brightest stars distance.}
%% General table references marker
\tablerefs{$^{a}$ - SIMBAD
$^{b}$ - \citet{sharina1999}
$^{c}$ - \citet{dalcanton1997}
}
\end{deluxetable*}

%% file: m64.tex
\begin{deluxetable*}{ccccccccccccc}
\rotate
\tablecaption{M64 Dwarf Sample\label{tab:m64}}

\tablewidth{\textwidth}

\tablehead{\colhead{Name} & \colhead{$\alpha$} & \colhead{$\delta$} & \colhead{Type} & \colhead{$m_g$} & \colhead{$m_i$} & \colhead{$g-i$} & \colhead{$r_e$} & \colhead{$\mu_{e,i}$} & \colhead{Star Mask}   & \colhead{D$_{\rm TRGB}$} & \colhead{$cz$} & \colhead{Other Names} \\ 
\colhead{} & \colhead{} & \colhead{} & \colhead{} & \colhead{(mag)} & \colhead{(mag)} & \colhead{} & \colhead{$\prime\prime$} & \colhead{(mag $\prime\prime^{-2}$)} & \colhead{Y/N} & \colhead{(Mpc)}  & \colhead{(km/s)}  & \colhead{} } 

\startdata
dw1255+2130 & 12:55:33.6 & 21:30:35.3 & dE & 20.37$\pm$0.06 & 19.83$\pm$0.06 & 0.54$\pm$0.03 & 5.3$\pm$0.4 & 25.55$\pm$0.1 & N & & &   \\
\enddata
\tablecomments{Photometric Properties of the dwarf satellite candidates found in the field around M64. }
%% General table references marker
\tablerefs{
}
\end{deluxetable*}

%% file: m104.tex
\begin{deluxetable*}{ccccccccccccc}
\rotate
\tablecaption{M104 Dwarf Sample\label{tab:m104}}

\tablewidth{\textwidth}

\tablehead{\colhead{Name} & \colhead{$\alpha$} & \colhead{$\delta$} & \colhead{Type} & \colhead{$m_g$} & \colhead{$m_i$} & \colhead{$g-i$} & \colhead{$r_e$} & \colhead{$\mu_{e,i}$} & \colhead{Star Mask}   & \colhead{D$_{\rm TRGB}$} & \colhead{$cz$} & \colhead{Other Names} \\ 
\colhead{} & \colhead{} & \colhead{} & \colhead{} & \colhead{(mag)} & \colhead{(mag)} & \colhead{} & \colhead{$\prime\prime$} & \colhead{(mag $\prime\prime^{-2}$)} & \colhead{Y/N} & \colhead{(Mpc)}  & \colhead{(km/s)}  & \colhead{} } 

\startdata
dw1237-1125 & 12:37:11.6 & -11:25:59.3 & dE & 18.28$\pm$0.06 & 17.28$\pm$0.24 & 0.99$\pm$0.23 & 10.0$\pm$0.4 & 24.59$\pm$0.24 & N & & &   \\
dw1237-1110 & 12:37:42.0 & -11:10:08.5 & dI & 18.84$\pm$0.11 & 18.21$\pm$0.09 & 0.63$\pm$0.03 & 5.8$\pm$0.6 & 24.04$\pm$0.17 & N & & &   \\
dw1238-1208 & 12:38:22.2 & -12:08:08.1 & dE & 22.63$\pm$0.15 & 22.02$\pm$0.22 & 0.61$\pm$0.37 & 3.4$\pm$0.2 & 27.04$\pm$0.31 & N & & &   \\
dw1238-1116 & 12:38:31.1 & -11:16:25.5 & dE & 21.07$\pm$0.34 & 20.67$\pm$0.36 & 0.4$\pm$0.2 & 5.8$\pm$1.1 & 26.6$\pm$0.27 & N & & &   \\
dw1238-1122 & 12:38:33.7 & -11:22:05.0 & dE & 16.5$^{**}$ & 15.6$^{**}$ & 0.89$^{**}$ & 34.7$^{**}$ & 26.2$^{**}$ & Y & & &   \\
dw1238-1102 & 12:38:58.3 & -11:02:09.6 & dE & 20.73$\pm$0.25 & 20.23$\pm$0.28 & 0.5$\pm$0.16 & 5.2$\pm$0.6 & 26.16$\pm$0.15 & N & & &   \\
dw1239-1152 & 12:39:09.0 & -11:52:36.6 & dE & 21.81$\pm$0.22 & 21.33$\pm$0.23 & 0.49$\pm$0.1 & 4.9$\pm$0.7 & 26.37$\pm$0.19 & N & & &   \\
dw1239-1159 & 12:39:09.1 & -11:59:12.2 & dE & 18.9$\pm$0.22 & 18.39$\pm$0.22 & 0.52$\pm$0.08 & 14.1$\pm$2.5 & 26.67$\pm$0.24 & N & & &   \\
dw1239-1143 & 12:39:15.3 & -11:43:08.1 & dE,N & 16.52$\pm$0.03 & 15.71$\pm$0.02 & 0.8$\pm$0.01 & 12.5$\pm$0.3 & 23.37$\pm$0.03 & N & & & NGC 4594 DW1$^b$ \\
dw1239-1154 & 12:39:22.4 & -11:54:23.9 & dE & 21.43$\pm$0.45 & 20.7$\pm$0.4 & 0.73$\pm$0.18 & 8.3$\pm$1.3 & 27.18$\pm$0.26 & N & & &   \\
dw1239-1113 & 12:39:32.7 & -11:13:36.0 & dE & 18.0$\pm$0.27 & 17.16$\pm$0.28 & 0.84$\pm$0.13 & 17.3$\pm$2.6 & 25.89$\pm$0.19 & N & & & NGC 4594 DGSAT-3$^a$ \\
dw1239-1118 & 12:39:37.4 & -11:18:33.1 & dE & 21.59$\pm$0.13 & 21.01$\pm$0.15 & 0.58$\pm$0.1 & 4.0$\pm$0.4 & 26.25$\pm$0.19 & N & & &   \\
dw1239-1106 & 12:39:41.9 & -11:06:00.4 & dE & 20.88$\pm$0.18 & 20.27$\pm$0.17 & 0.61$\pm$0.09 & 5.9$\pm$0.9 & 26.18$\pm$0.18 & N & & &   \\
dw1239-1120 & 12:39:51.5 & -11:20:28.7 & dE & 19.41$\pm$0.1 & 18.82$\pm$0.1 & 0.59$\pm$0.03 & 7.0$\pm$0.6 & 25.43$\pm$0.1 & N & & & NGC 4594 DGSAT-2$^a$ \\
dw1239-1144 & 12:39:54.9 & -11:44:45.5 & dE & 17.33$\pm$0.29 & 16.63$\pm$0.26 & 0.7$\pm$0.14 & 22.4$\pm$4.5 & 25.56$\pm$0.13 & N & & & NGC 4594 DGSAT-1$^a$ \\
dw1240-1118 & 12:40:09.4 & -11:18:49.8 & dE,N & 15.9$\pm$0.03 & 15.09$\pm$0.03 & 0.81$\pm$0.01 & 15.1$\pm$0.4 & 23.58$\pm$0.04 & N & & &   \\
dw1240-1140 & 12:40:17.6 & -11:40:45.7 & dE,N & 19.26$\pm$0.46 & 18.32$\pm$0.48 & 0.94$\pm$0.06 & 13.1$\pm$2.8 & 26.42$\pm$0.17 & N & & &   \\
dw1240-1155 & 12:40:59.5 & -11:55:48.0 & dI & 18.76$\pm$0.04 & 18.58$\pm$0.05 & 0.17$\pm$0.02 & 4.7$\pm$0.2 & 23.54$\pm$0.06 & N & & &   \\
dw1241-1131 & 12:41:02.8 & -11:31:43.7 & dE,N & 19.78$\pm$0.17 & 18.97$\pm$0.24 & 0.8$\pm$0.11 & 9.1$\pm$0.9 & 26.12$\pm$0.15 & Y & & &   \\
dw1241-1210 & 12:41:03.2 & -12:10:46.6 & dI & 19.41$\pm$0.05 & 19.36$\pm$0.05 & 0.06$\pm$0.04 & 4.2$\pm$0.2 & 24.99$\pm$0.1 & N & & &   \\
dw1241-1123 & 12:41:09.5 & -11:23:55.3 & dE? & 21.08$\pm$0.38 & 20.28$\pm$0.36 & 0.8$\pm$0.09 & 12.2$\pm$2.5 & 28.34$\pm$0.19 & N & & &   \\
dw1241-1105 & 12:41:10.2 & -11:05:50.0 & dE & 21.89$\pm$0.11 & 20.9$\pm$0.14 & 0.99$\pm$0.06 & 2.3$\pm$0.2 & 25.16$\pm$0.14 & N & & &   \\
dw1241-1153 & 12:41:12.1 & -11:53:29.7 & dE,N & 18.33$\pm$0.22 & 17.61$\pm$0.19 & 0.72$\pm$0.06 & 15.3$\pm$2.3 & 25.82$\pm$0.17 & N & & &   \\
dw1241-1155 & 12:41:18.7 & -11:55:30.8 & dE,N & 17.48$\pm$0.11 & 16.73$\pm$0.13 & 0.75$\pm$0.06 & 17.0$\pm$1.2 & 25.12$\pm$0.1 & N & & &   \\
dw1242-1116 & 12:42:43.8 & -11:16:26.0 & dE,N & 18.09$\pm$0.22 & 17.51$\pm$0.17 & 0.58$\pm$0.07 & 24.4$\pm$4.6 & 26.67$\pm$0.21 & N & & &   \\
dw1242-1129 & 12:42:49.6 & -11:29:21.5 & dE/dI & 21.08$\pm$0.23 & 20.35$\pm$0.24 & 0.73$\pm$0.13 & 3.3$\pm$0.5 & 25.21$\pm$0.24 & N & & &   \\
dw1243-1137 & 12:43:18.0 & -11:37:33.0 & dE & 21.56$\pm$0.19 & 20.53$\pm$0.13 & 1.03$\pm$0.07 & 4.4$\pm$0.8 & 25.86$\pm$0.29 & N & & &   \\
\enddata
\tablecomments{Photometric Properties of the dwarf satellite candidates found in the field around M104. Galaxies marked with $^{**}$ were either very non-S\'{e}rsic or had some other issue with the fitting and the photometry should be treated with caution.}
%% General table references marker
\tablerefs{$^{a}$ - \citet{javanmardi_m101}
$^{b}$ - \citet{caldwell83}
}
\end{deluxetable*}